\documentclass[aps,pra,twocolumn,reprint,nofootinbib,floatfix]{revtex4}
\usepackage{graphicx,bm,amsmath}
\usepackage[pdftex]{color} 
\usepackage{hyperref}
\usepackage{color}
\usepackage{ulem} 
\usepackage[paperwidth=8.5in,paperheight=11in,centering,hmargin=2cm,vmargin=2.5cm]{geometry} 

\def\ket#1{|\, #1\,\rangle}
\def\bra#1{\langle\,#1\,|}

\begin{document}

\title{Non-Abelian braiding of light}

\author{Thomas~Iadecola} \affiliation{Physics Department, Boston
  University, Boston, Massachusetts 02215, USA}
  
\author{Thomas~Schuster} \affiliation{Physics Department, Boston
 University, Boston, Massachusetts 02215, USA}

\author{Claudio~Chamon} \affiliation{Physics Department, Boston
  University, Boston, Massachusetts 02215, USA}
  
\date{\today}

\begin{abstract}
Many topological phenomena first proposed and observed in the context
of electrons in solids have recently found counterparts in photonic and acoustic
systems. In this work, we demonstrate that non-Abelian Berry phases
can arise when coherent states of light are injected into
``topological guided modes" in specially-fabricated photonic waveguide
arrays. These modes are photonic analogues of topological zero modes
in electronic systems. Light traveling inside spatially well-separated
topological guided modes can be braided, leading to the accumulation
of non-Abelian phases, which depend on the order in which the guided beams
are wound around one another. Notably, these effects survive the
limit of large photon occupation, and can thus also be understood as wave
phenomena arising directly from Maxwell's equations, without
resorting to the quantization of light. We propose an optical interference
experiment as a direct probe of this non-Abelian braiding of light.
\end{abstract}

\maketitle

Manifestations of topology in physical systems, specifically in the form of so-called geometric phases,~\cite{Pancharatnam,Berry1} have risen to prominence over the last three decades.  Geometric phases were shown by M.V.~Berry to arise in quantum systems under cyclic adiabatic variation of parameters,~\cite{Berry1} but it was later understood~\cite{Berry2} that this phase had been discovered thirty years earlier in the context of classical optics by S.~Pancharatnam.~\cite{Pancharatnam}  The close analogy between quantum mechanics and classical optics has remained over the years, and many of the striking topological states of matter associated with electrons in solids, such as topological insulators and semimetals,~\cite{HasanKaneReview} have recently found counterparts in photonic~\cite{Peleg,HaldaneRaghu,RaghuHaldane,Wang,Khanikaev,Zilberberg,RechtsmanFloquet,Lu,LuReview} and acoustic~\cite{ProdanProdan,Berg,KaneLubensky,Po,Susstrunk,WangPhonon} systems.  The present work aims to further highlight and deepen the connection between topological phenomena in solids and in classical wave mechanics by demonstrating a new facet to this correspondence. We demonstrate the existence of a classical analogue of the \textit{non-Abelian} Berry phase~\cite{WilczekZee} that arises from ``braiding" topological defects in solids.

One route to non-Abelian Berry phases in electronic systems lies in the remarkable physics of zero modes.  As was pointed out in pioneering work by R.~Jackiw with C.~Rebbi~\cite{JackiwRebbi} and P.~Rossi,~\cite{JackiwRossi} localized zero-energy fermionic states can bind to topological defects in an order parameter, such as kinks in one dimension and vortices in two dimensions.  In systems where electric charge is a good quantum number, these zero modes carry charges that are fractions of the ``fundamental" electron charge.~\cite{JackiwRebbi,SuSchriefferHeeger,Hou}  In chiral superconductors, where charge conservation is broken, these localized modes are Majorana bound states with non-Abelian statistics.~\cite{JackiwRossi,ReadGreen,Ivanov,Kitaev} At the mean-field level, where interactions between electrons are neglected, this non-Abelian statistcs can be understood in terms of the accumulation of non-Abelian Berry phases as defects are adiabatically exchanged with one another.~\cite{Stern}  The non-Abelian nature of this process manifests itself in dependence of these phases on the order in which the defects are interchanged, in contrast to the usual ``Abelian" Berry phase.

In this Letter, we propose a novel means of realizing topological zero modes in photonic, rather than electronic, systems, and demonstrating their non-Abelian braiding directly. The proposed realization consists of non-interacting photons propagating in the non-trivial background of a photonic lattice with topological defects whose positions are controllable. Light channeled into ``topological guided modes" localized at these defects can be braided, leading to the accumulation of non-Abelian phases that depend on the order in which the braids occur.  We demonstrate that this effect manifests itself at both the quantum and classical levels, owing to the linearity of the equations of motion for noninteracting light.

The zero modes we propose to realize are photonic analogues of Kekul\'e zero modes in graphene,~\cite{Hou} which are bound to vortices in the complex order parameter $\Delta(\bm r)$ describing a dimerization pattern in the hexagonal lattice (Fig.~1).  The translation between electrons and photons is achieved by replacing the sites of the lattice with waveguides embedded in a bulk optical medium (\textit{e.g.}~fused silica),~\cite{SzameitReview} which are extended in the $z$ direction and whose $x$-$y$ positions mimic the 2D positions of the carbon atoms in the distorted graphene lattice.  The wave equation for the paraxial propagation of light in such a waveguide array maps directly onto the time-dependent Schr\"odinger equation (SE), where the time coordinate $t$ in the SE is replaced by the coordinate $z$ along the direction of light propagation.  This wave equation can be further mapped, using coupled mode theory,~\cite{LedererReview} to a linear differential equation that is in one-to-one correspondence with the noninteracting tight-binding model of the electronic system.    The waveguides themselves can therefore be thought of as the world lines of the carbon atoms, with straight waveguides corresponding to a lattice that is static in time.  

Waveguide arrays of this type have been realized, written into bulk materials with exquisite precision using femtosecond lasers,~\cite{Davis} (see Ref.~\onlinecite{SzameitReview} for a review).  They have been used to mimic graphene and other electronic systems with both static lattices~\cite{Peleg,Bahat-Treidel,RechtsmanStrain,Mukherjee} and ones that change as a function of time.~\cite{RechtsmanFloquet,Zilberberg}  The experimental protocol suggested here hinges on this capability to execute a braiding procedure in which three vortices are wound around one another.  As we show below, a protocol can be chosen that reveals the non-Abelian nature of the braiding directly via interference between different braiding patterns.

\begin{figure}
\centering
a)\includegraphics[width=.215\textwidth,page=1]{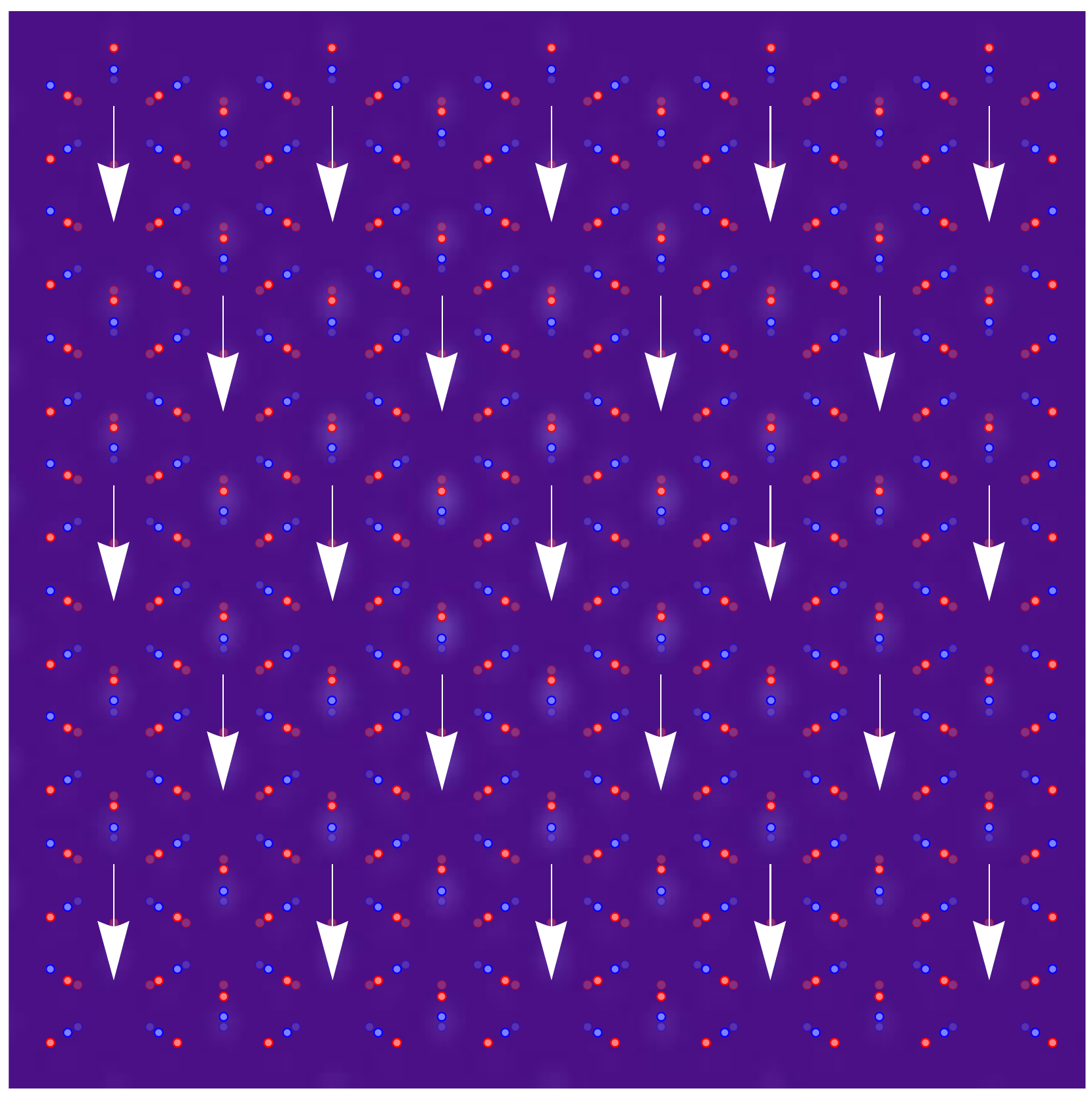}
\hspace{2mm} 
b)\includegraphics[width=.21\textwidth,page=2]{Figures}
\caption{a) Uniform Kekul\'e distortion in a hexagonal waveguide lattice at a slice of constant $z$.  The faint circles represent the original hexagonal lattice, while darker circles represent the distorted lattice.  Sublattice $A$ is colored red, while sublattice $B$ is colored blue.  The overlaid vector field represents the magnitude and direction of the order parameter $\Delta(\bm r)$. The background is an image of a low-lying eigenmode of Eq.~(\ref{waveguide equation}).  The intensity pattern represents the amplitude of the electric field in each waveguide.  b) Waveguide lattice in the presence of a vortex in the Kekul\'e pattern, with an overlaid order-parameter vector field showing circulation around the vortex core.  The background is an intensity plot of the localized zero-mode wavefunction.  We have chosen a vortex profile in which only sublattice $A$ is displaced. \label{fig: lattice}}
\end{figure}

Our starting point is the coupled-mode equation for paraxial light propagation through the waveguide array,
\begin{equation}\label{waveguide equation}
i\, \partial_z\, \psi_{\bm r}(z) = \sum^3_{j=1} H_{\bm r,\bm r\pm\bm s_j}\, \psi_{\bm r\pm\bm s_j}(z).
\end{equation}
Here, the vector $\bm r$ is defined on a hexagonal lattice divided into two triangular sublattices that we call $A$ and $B$. The vectors $+\bm s_j$ ($j$=1,2,3) connect the point $\bm r$ in sublattice $A$ to its three nearest neighbors in sublattice $B$, located a distance $a$ away, and $-\bm s_j$ connect points in $B$ to their neighbors in $A$.  For simplicity, we assume the light in the array to be monochromatic, so that Eq.~(\ref{waveguide equation}) is frequency-independent.  Eq.~(\ref{waveguide equation}) is formally identical to a noninteracting tight-binding model in which the time coordinate $t$ has been replaced by the longitudinal coordinate $z$. Its solutions can be decomposed into eigenmodes $\psi_{\nu,\bm r}(z)=\psi_{\nu,\bm r}\, e^{-i\kappa_\nu z}$, where $\nu=0,\dots,2N-1$, with $N$ the number of unit cells in the waveguide array, labels the ``energy eigenvalues" $\kappa_\nu$.  The ``wavefunction" $\psi_{\bm r}(z)$ in Eq.~(\ref{waveguide equation}) is related to the amplitude of the local electric field $\bm E_{\bm r}(z,t)$ on the ``site" $\bm r$.  It is useful to work with the \textit{quantized} electric field operator\cite{VogelWelsch}
\begin{equation}\label{E field}
\hat{\bm E}_{\bm r}(z,t) = \sum_{\nu}\psi_{\nu,\bm r}\, e^{i[(k_\omega-\kappa_\nu)z-\omega t]}\, \hat{b}^{\phantom\dagger}_{\nu}\, \bm n+\text{H.c.}\, 
\end{equation}
where $\bm n$ is a unit vector in the $x$-$y$ plane describing the polarization of the field and $k_\omega$ satisfies the dispersion relation of a single waveguide.  The ladder operators $\hat{b}_{\nu}$ satisfy the bosonic commutation relations $[\hat b^{\phantom\dagger}_{\nu},\hat b^\dagger_{\nu^\prime}]=\delta_{\nu,\nu^\prime}$.  [See Supplemental Material (SM) for details on the quantization procedure.]

The ``Hamiltonian" in Eq.~(\ref{waveguide equation}) depends exponentially on distances between nearest-neighbor waveguides (we neglect longer-range couplings for the moment).  We take $H_{\bm r,\bm r+\bm s_j}=-t-\delta t_{\bm r,j}$, where $\delta t_{\bm r,j}=\Delta(\bm r)\, e^{i\bm K_+\cdot\bm s_j}\, e^{2i\bm K_+\cdot\bm r}/2+\text{c.c.}$ (we denote by $\bm K_{\pm}$ the locations of the two inequivalent Dirac points, at opposite corners of the hexagonal Brillouin zone).  Here, the parameter $t$ describes the evanescent couplings of the waveguides, and the position-dependent function $\delta t_{\bm r, j}$ describes modulations of these couplings due to displacements of the waveguides from their original $x$-$y$ positions.  The complex-valued Kekul\'e order parameter $\Delta(\bm r)$ controls the distortion of the lattice.~\cite{Hou}  A vortex in $\Delta(\bm r)$ is a defect in this distortion pattern, but not the lattice itself.  The order parameter in the presence of a single vortex centered at the origin reads $\Delta(\bm r) = \Delta_0(r)\, e^{i(\alpha-\theta)}$ in polar coordinates $\bm r = (r,\theta)$.  Here, $\Delta_0(r)=\Delta\, \tanh(r/\ell_0)$ describes the spatial profile of the vortex, which has a core radius $\ell_0$, and $\alpha$ is the phase of the order parameter.  

The zero mode in the presence of this vortex profile can be found by setting the left-hand side of Eq.~(\ref{waveguide equation}) to zero and solving for $\psi_{\bm r}$. The zero-mode solution is tightly localized near the core of the vortex, with a size of order $1/\Delta$, and has support on sublattice $A$ only [see Fig.~\ref{fig: lattice}(B)].  [If we send $\theta\to-\theta$ in $\Delta(\bm r)$, the zero mode has support on sublattice $B$ instead.~\cite{Hou}]  This means that light propagating in the zero mode travels as if confined to an ``optical fiber" located at the vortex core, albeit with evanescent decay into neighboring waveguides in the same sublattice.  However, the zero mode differs crucially from a mode in an optical fiber, both because it takes the distortion of an entire waveguide lattice to create, and because it depends on the topological nature of this distortion.  These ``topological guided modes" are responsible for the non-Abelian effects described below.

The above discussion neglects the presence of various lattice effects that can shift the zero mode eigenvalue to a finite momentum.  In principle, this makes possible the accumulation of path-dependent (\textit{i.e.}~non-topological) dynamical phases during braiding.  However, as we show in detail in the SM, these effects can be neglected to resonable precision without fine-tuning.  There, we estimate (conservatively) that we can perform thousands of braids before the effects of dynamical phases begin to manifest themselves.

To facilitate our discussion of the non-Abelian braiding of vortices like the one above, we work in the continuum limit, where the analytical form of the vortex wavefunction is known.  (Exact diagonalization calculations corroborating these results are reviewed in the SM.)  On length scales much longer than $a$, the operator that annihilates a photon in the zero mode is
\begin{equation}\label{vortex operator}
\hat b^{\phantom\dagger}_0=\int \mathrm d \bm r
  \left[
    u({\bm r})\,\hat b^{\phantom\dagger}_{+}({\bm r})
    +
    u^*({\bm r})\,\hat b^{\phantom\dagger}_{-}({\bm r})
    \right]
  \;,
\end{equation}
where $\hat b_{\pm}(\bm r)$ annihilate a photon at the Dirac points $\bm K_{\pm}$.  (These operators are assumed to be normalized such that the canonical commutation relation $[\hat b^{\phantom\dagger}_0,\hat b^\dagger_0]=1$ holds.)  The function $u(\bm r)\equiv u(r)$ is given, up to normalization, by~\cite{Hou}
\begin{equation}\label{wavefunction}
u(r)=e^{i(\alpha/2+\pi/4)}\, e^{-\int_0^r\mathrm dr^\prime\, \Delta_0(r^\prime)}.
\end{equation}
Note that the zero mode amplitude $u(r)$ is \textit{double-valued}: when $\alpha$, the phase of the order parameter $\Delta(\bm r)$, changes by $2\pi$, $u(r)$ acquires a minus sign.  The origin of this double-valuedness lies in the fact that $u(r)$ is a solution to the Dirac equation, and therefore must transform as a spinor under changes in $\alpha$.

The quantum states created by $\hat b^\dagger_0$ can be connected to the macroscopic state of light in the waveguide array by defining the coherent state
\begin{equation}\label{coherent state def}
|\lambda\rangle=e^{-|\lambda|^2/2}\, e^{\lambda\hat b^{\dagger}_0}\, |0\rangle
\end{equation}
with mean photon number $\langle\hat b^{\dagger}_0\hat b^{\phantom\dagger}_0\rangle=|\lambda|^2$.  These states provide a faithful description of the electric field in the waveguide array in the large-$|\lambda|$ limit when the input light is a coherent, monochromatic laser spot centered on the vortex core.  In this case, the contribution of the photonic zero mode to the classical electric field is given by [c.f.~Eq.~\eqref{E field}]
\begin{align}\label{classical E field}
\bm E_{0,\bm r}(z,t) = \lambda\, u(r)\, e^{i(\bm K_{+}\cdot\bm r + k_\omega\, z-\omega t)}\, \bm n+\text{c.c.},\, 
\end{align}
with $u(r)$ given in Eq.~\eqref{wavefunction} and $\lambda$ defined by Eq.~\eqref{coherent state def}.  The ease with which one can translate between the quantum and classical descriptions of the waveguide array owes to the fact that photons in the array are noninteracting.  Indeed, an alternative way of deriving Eq.~\eqref{classical E field} is to solve Maxwell's equations directly in the presence of a vortex.

\begin{figure}
\centering
\includegraphics[width=.475\textwidth, page=3]{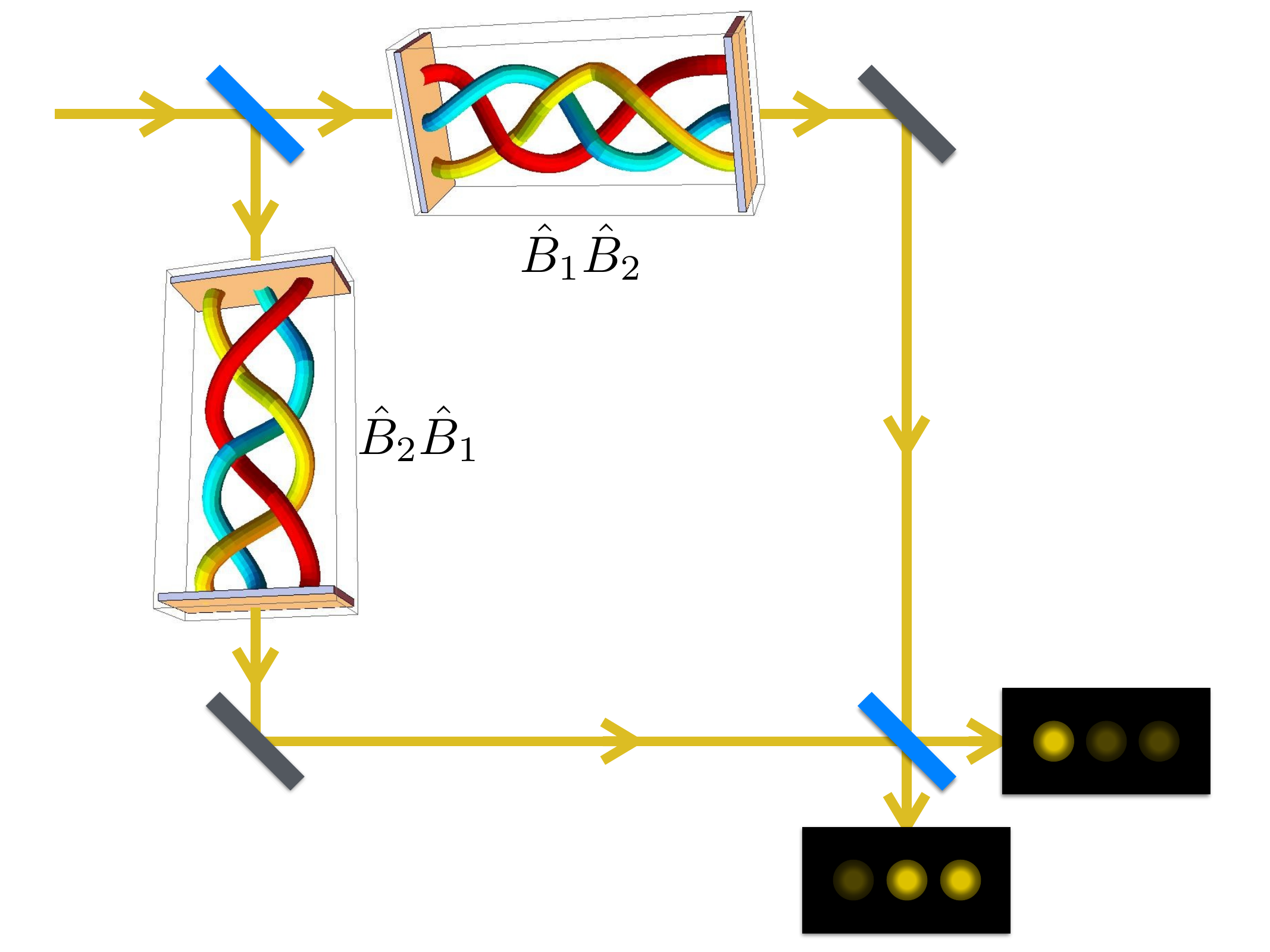}
\caption{Schematic of the proposed photonic non-Abelian interferometer.  A coherent laser beam (one of three, one for each vortex in a lattice) passes through a 50/50 beam splitter (blue) and simultaneously enters two separate photonic lattices fabricated with the two braids to be compared.  The two output beams are reflected by mirrors (grey) into another beam splitter that interferes the two signals and outputs the sum and difference to separate screens (black).  For the choice of braids used here, one screen shows a bright and two dark spots, while the other screen shows the ``logical complement" of the first, with one dark and two bright spots.  \label{fig: PNAI}}
\end{figure}

Let us now study the braiding of zero modes in an infinite system with $v$ vortices at $z$\textit{-dependent} positions $\bm R_i(z)$ ($i=1,\dots,v$).  The order parameter in the presence of this vortex configuration is given by
\begin{equation}
\Delta(\bm{r}) = \Delta \prod_{j=1}^{n} \tanh(|\bm{r}-\bm{R}_j|/\ell_0)\, e^{i [\alpha_j - \text{arg}(\bm{r}-\bm{R}_j)]}\, ,
\end{equation}
and the zero-mode Hilbert space is spanned by the operators $\hat b^{\dagger}_{0,i}$.  A clockwise adiabatic exchange of vortices $i$ and $i+1$ is implemented by winding the vortex-core coordinates $\bm R_i(z)$ and $\bm R_{i+1}(z)$ around one another as functions of increasing $z$ (see, e.g., the braids in Fig.~\ref{fig: PNAI}).  In the SM, we show that the nontrivial nature of this winding process stems from the double-valuedness of $u(r)$ under $\alpha\to\alpha-2\pi$. (We verified this statement for the zero-mode wavefunction on the lattice via a numerical tight-binding calculation, as we describe in the SM.)  The effect of this exchange, up to a gauge choice, is to map $\hat b^{\dagger}_{0,i}\to\hat b^{\dagger}_{0,i+1}$ and $\hat b^{\dagger}_{0,i+1}\to -\hat b^{\dagger}_{0,i}$ while leaving all other vortex operators unchanged, similarly to the case of Majorana~\cite{Ivanov} and Dirac zero modes~\cite{Yasui} in electronic systems.  The operator that implements this exchange is
\begin{equation}\label{braiding generator}
\hat{\tau}_i = e^{\pi\,(\hat{b}_{0,i+1}^\dagger\hat{b}^{\phantom\dagger}_{0,i}-\hat{b}_{0,i}^\dagger\hat{b}^{\phantom\dagger}_{0,i+1})/2}.
\end{equation}
One verifies by direct calculation that these operators satisfy $[\hat{\tau}_i,\hat{\tau}_j]=0$ for $|i-j|>1$ and $\hat{\tau}_i\hat{\tau}_j\hat{\tau}_i=\hat{\tau}_j\hat{\tau}_i\hat{\tau}_j$ for $|i-j|=1$, and therefore form a unitary representation of $\mathcal B_{v}$, the braid group on $v$ strands.~\cite{NayakReview}  The action of these generators on a state
\begin{equation}
|n_1,\dots,n_v\rangle=\prod^{v}_{i=1}\frac{(\hat{b}^\dagger_{0,i})^{n_i}}{\sqrt{n_i!}}\, |0\rangle,
\end{equation}
representing a system with $v$ vortices, each with a fixed number of photons, is given by
\begin{equation}\label{fixed occ braid}
\hat \tau_i\, |\!\dots,n_i,n_{i+1},\dots\rangle \!=\! (-1)^{n_{i+1}} |\!\dots,n_{i+1},n_i,\dots\rangle.
\end{equation}

We now explain the consequences of the above phase factor, which arises purely from braiding the vortices. Consider the case of two vortices, and the operation of winding one around the other, which restores them to their initial locations. We begin by considering i) an example where this phase factor does not lead to unitary operations on a multi-dimensional space of degenerate states, and then turn to ii) an example where it does. For case i), consider any eigenstate of occupation number, $|n_1,n_2\rangle$, which upon winding goes to the state $(-1)^{n_1+n_2}\,|n_1,n_2\rangle$; clearly, the initial and final states are equal up to a phase, and the braiding operation acts on a one-dimensional space only. In contrast, consider case ii), where we take an initial state that is not an eigenstate of occupation number, but rather a superposition of states with occupation 0 and 1, for example.  Then winding takes
\begin{equation}
\frac{|0\rangle\!+\!|1\rangle}{\sqrt{2}}
  \otimes
  \frac{|0\rangle\!+\!|1\rangle}{\sqrt{2}}
\longrightarrow
  \frac{|0\rangle\!-\!|1\rangle}{\sqrt{2}}
  \otimes
  \frac{|0\rangle\!-\!|1\rangle}{\sqrt{2}}
  \;.
\end{equation}
Clearly, the initial and final states are different (here, they are orthogonal).  This is a crucial hallmark of the non-Abelian nature of the action of the braiding generators on these states.
Thus, the essential ingredient necessary for braiding to connect states in a multi-dimensional Hilbert space is that the initial state of the zero mode system consists of superpositions of states with even and odd numbers of photons.

Such superpositions do not only exist within the domain of quantum optics.  The coherent states defined in Eq.~\eqref{coherent state def} consist of a superposition of photon states with \textit{all} occupation numbers, and can be created by shining a coherent laser beam centered on a single vortex core.  Let us now consider a system of $v$ vortices, into each of which is loaded a coherent state of photons:
\begin{equation}
|\lambda_1,\dots,\lambda_{v}\rangle=\prod^{v}_{i=1}e^{-|\lambda_i|^2/2}\, e^{\lambda_i\hat b^{\dagger}_{0,i}}\, |0\rangle.
\end{equation}
The action of the braiding generators on these states is found from Eq.~\eqref{fixed occ braid} to be
\begin{equation}
\hat{\tau}_i|\dots,\lambda_i,\lambda_{i+1},\dots\rangle=|\dots,-\lambda_{i+1},\lambda_i,\dots\rangle.
\end{equation}
To see that this braiding operation connects quantum states in a multi-dimensional space of degenerate states, consider again the winding of two vortices, which restores their initial locations. For a coherent state, this operation takes
\begin{equation}\label{2-vortex winding coherent}
  |\lambda_1,\lambda_2\rangle
  \longrightarrow
  |-\lambda_1,-\lambda_2\rangle
  \;.
\end{equation}
The overlap of these two states is $\langle \lambda_1,\lambda_2 |-\lambda_1,-\lambda_2\rangle=e^{-2(|\lambda_1|^2+|\lambda_2|^2)}$, whose magnitude is smaller than one, indicating that the states are linearly independent. This demonstrates that the Hilbert space spanned by the braiding operations is multi-dimensional. Moreover, in the limit where a large number of photons are loaded into the zero modes (large $|\lambda_{1,2}|$, which is to be expected from a laser), the overlap is exponentially small, and the initial and final states become orthogonal.

At this point, it is worthwhile to reflect on the significance of the fact that the effect of braiding the vortices manifests itself at the level of coherent states, which are essentially classical.  While the braiding of these vortices can be understood at the quantum (\textit{i.e.}~few-photon) level, its effects permeate the entire zero-mode Hilbert space for arbitrary occupation numbers.  Consequently, the quantum action of the braiding generators survives the limit of large occupations, so that macroscopic effects of this braiding can be seen. In particular, observe that $\lambda_i\to-\lambda_i$ under a braid corresponds, in the limit of large $|\lambda_i|$, to a change in the sign of the electric field $\bm E_i$ near the core of the $i$th vortex [c.f.~Eq.~\eqref{classical E field}]. This observation forms the basis of our discussion below.

We now propose an experiment that could provide direct and unambiguous evidence for non-Abelian braiding at the level of coherent states.  The photonic non-Abelian interferometer (PNAI) that we propose, shown in Fig.~\ref{fig: PNAI}, consists of two separate photonic lattices, each with three vortices, with waveguides written into the host medium in such a way that the vortex cores wind adiabatically around one another according to a specific pattern.  (The adiabatic condition here corresponds to demanding that any change in the position of each waveguide as a function of $z$ occur on length scales much larger than the inverse photonic bandgap.)  In one lattice, the waveguide pattern executes a braid $\hat{B}_1\hat{B}_2$, while in the other, the waveguide pattern implements the braid $\hat{B}_2\hat{B}_1$.  Interfering the light output from the two lattices reveals the defining feature of non-Abelian braiding, namely that performing the same braids in different orders yields different results.  We now proceed through each stage of the interferometer setup.

We start with the input stage of the PNAI.  We assume that the input light comes from three monochromatic, coherent laser sources, each focused on the core of a single vortex so that there is large overlap with the zero modes.  Each of the three input beams is split by a 50/50 dielectric beam splitter, so that the light entering each vortex core comes from the same source beam.  We will denote the light entering the upper branch of the interferometer in Fig.~\ref{fig: PNAI} as $|\lambda_1,\lambda_2,\lambda_3\rangle$, and $|\tilde\lambda_1,\tilde\lambda_2,\tilde\lambda_3\rangle$ for the lower branch.  (The first beam splitter enforces the phase relation $\tilde\lambda_j=-i\,\lambda_j$.)

For the braiding stage of the interferometer, we choose, for example, $\hat{B}_1=\hat \tau_2\hat \tau_1$ and $\hat{B}_{2}=\hat\tau_2\hat\tau^{-1}_1\hat{\tau}_2\hat{\tau}_1^{-1}$ (see Fig.~\ref{fig: PNAI}), which ensures that the vortices on the output facets of both lattices are in the same order.~\cite{Footnote}  In the SM, we provide more information about how to write these braiding patterns into the waveguide array.  The output states from the two braids are
\begin{equation}\label{out B1B2}
\hat B_1\hat B_2\, |\lambda_1,\lambda_2,\lambda_3\rangle = |-\lambda_1,\lambda_2,-\lambda_3\rangle
\end{equation}
and
\begin{equation}\label{out B2B1}
\hat B_2\hat B_1\, |\tilde\lambda_1,\tilde\lambda_2,\tilde\lambda_3\rangle = |-\tilde\lambda_1,-\tilde\lambda_2,\tilde\lambda_3\rangle.
\end{equation}


In the final stage of the interferometer, the output beams from the braiding stage are combined at another beam splitter.  
The sign differences between the coherent states exiting the two branches of the interferometer cause the light to interfere constructively at one detector and destructively at the other.  Which detector this is for each vortex depends on the relative signs of $\lambda_i$ and $\tilde\lambda_i$, which in turn depend on the braids (see Fig.~\ref{fig: PNAI}, and SM for more details).  Since the effects of dynamical phases can be heavily suppressed in a controlled manner, as discussed earlier, the only source of this interference is the noncommutativity of braiding the vortices.

In summary, we have demonstrated in this Letter a means to realize photonic analogues of topological zero modes in photonic lattices.  We demonstrated that these ``topological guided modes" can be understood at both the quantum and classical levels when the photons in the waveguide array are weakly interacting.  We further proposed a photonic non-Abelian interferometer, feasible with current technology, to detect unambiguous signatures of the non-Abelian Berry phases that result from braiding these topological guided modes.

\begin{acknowledgments}
We thank Luca Dal Negro for inspiring discussions, and Chang-Yu Hou, Christopher
Mudry, Titus Neupert, and Luiz H. Santos for valuable feedback on the manuscript.  
 T.I. was supported by the National Science Foundation Graduate Research Fellowship Program
under Grant No.~DGE-1247312, and C.C. was
supported by DOE Grant DEF-06ER46316.
\end{acknowledgments}

\bibliographystyle{apsrev}

\bibliography{refs_braiding_light_quanta}

\begin{thebibliography}{38}
\bibitem[{Foo()}]{Footnote}
\bibinfo{howpublished}{Of course, this is not the only braid we could have
  chosen. For example, the choice $\hat{B}_1=\hat\tau_1\hat\tau_2$ and $\hat
  B_2=\hat\tau_2\hat\tau_1$ also satisfies the requirement that vortices end up
  in the same order in both braids.}
\end{thebibliography}

\appendix

\begin{widetext}

\hfill

\begin{center}
\textbf{\large Supplemental Material: Non-Abelian braiding of light}
\end{center}

\section{Quantized electromagnetism in a waveguide lattice}

We seek to quantize the electromagnetic field in the presence of a spatially varying relative permittivity $\epsilon(\bm{r})$, and follow the notation of $\textit{Quantum Optics}$, by Vogel and Welsch.~\cite{VogelWelsch} (We adopt the convention here that $\bm r$ is a three-dimensional vector when it enters the argument of a function, and a two-dimensional vector in the $x$-$y$ plane when it appears as a subscript.) We denote the vector potential by $\bm A(\bm{r},t)$, and work in the Coulomb gauge $\nabla \cdot \bm{A} = 0$.  In the absence of free charges we also have $A_0 = 0$. We work under the assumption that $\nabla \epsilon (\bm{r}) \cdot \bm{A}$ is small. With this the classical Hamiltonian is
\begin{equation}\label{classical Hamiltonian}
H  = \int \mathrm{d}^3r\,  \bigg(  \frac{1}{\epsilon(\bm{r})\epsilon_0}\,  \bm{\Pi}^2 +  \frac{1}{\mu_{0}} [\nabla \times \bm{A}]^2 \bigg)
\end{equation} 
with $\bm{\Pi} (\bm{r},t)$ the momentum conjugate to $\bm{A}(\bm{r},t)$:
\begin{equation}
\bm{\Pi} (\bm{r},t) =  \epsilon(\bm{r})\epsilon_0\,   \dot{\bm{A}}(\bm{r},t).\\
\end{equation} 
We then have the canonical Poisson brackets:
\begin{equation}
\{ \bm{A}(\bm{r},t), \bm{\Pi}(\bm{r}',t)  \} = \delta ( \bm{r} - \bm{r}').
\end{equation} 
We quantize Eq.~(\ref{classical Hamiltonian}) by promoting $\bm{A}$ and $\bm{\Pi}$ to operators obeying equal-time commutation relations
\begin{equation}
[ \hat{\bm{A}}(\bm{r},t), \hat{\bm{\Pi}}(\bm{r}',t)  ]  = i\hbar\, \delta ( \bm{r} - \bm{r}')
\end{equation} 
with all other commutators vanishing.  Our operator equations of motion,
\begin{equation}
\dot{\hat{\bm{A}}}(\bm{r},t) = \frac{1}{i\hbar}[ \hat{\bm{A}}(\bm{r},t), \hat{H}(t)  ] = \frac{1}{\epsilon (\bm{r})\epsilon_0} \hat{\bm{\Pi}}(\bm{r},t), \\
\end{equation}
\begin{equation}
\dot{\hat{\bm{\Pi}}}(\bm{r},t) = \frac{1}{i\hbar}[ \hat{\bm{\Pi}}(\bm{r},t), \hat{H}(t)  ] = \frac{1}{\mu_0} \nabla^2 \hat{\bm{A}}(\bm{r},t),
\end{equation} 
lead to the second-order equation of motion for $\hat{\bm{A}}(\bm{r},t)$ :
\begin{equation}
\nabla^2 \hat{\bm{A}} (\bm{r},t) - \frac{\epsilon(\bm{r})}{c^2} \frac{\partial^2 \hat{\bm{A}}(\bm{r},t)}{\partial t^2} = 0.
\end{equation} 
For a quasi-two-dimensional array of waveguides, our relative permittivity takes the form $\epsilon (x,y,z) = \epsilon_r [1 + \Delta \epsilon (x,y)]$. We can solve this using separation of variables with $\hat{\bm{A}} (\bm{r},t) \equiv \sum_{ \nu,\omega} \bm{A}_{\nu,\omega}(\bm{r})\, T_{\nu,\omega}(t)\,  \hat{b}_{\nu,\omega} + \text{H.c.}$, where we are anticipating two indices $\nu$ and $\omega$ which will label solutions of the resulting equations for $\bm{A}(\bm{r})$ and $T(t)$:
\begin{equation}
\ddot{T}_{\nu,\omega}(t) + \omega^2\, T_{\nu,\omega}(t)  = 0 \\
\end{equation}
and 
\begin{equation}\label{helmholtz}
\nabla^2 \bm{A}_{\nu,\omega} (x,y,z) + \frac{\omega^2 \epsilon_r}{c^2} \bm{A}_{\nu,\omega}(\bm{r})  + \frac{\omega^2  \epsilon_r \Delta \epsilon (x,y)}{c^2} \bm{A}_{\nu,\omega}(\bm{r}) = 0. \\
\end{equation}
The first equation is easily solved with $T_{\nu,\omega}(t) = T(0)e^{-i \omega t}$, and restricting $\omega > 0$. The second equation is the familiar Helmholtz equation, which the classical vector potential would also obey. We are interested in waves propagating primarily in the positive $z$-direction, and thus consider solutions of the form
\begin{equation}\label{Atopsi}
\bm{A}_{\nu,\omega}(x,y,z) = -i \omega^{-1}\,  \psi_{\nu,\omega}(x,y,z)\,   e^{i \frac{\omega}{v} z}\, \bm{n}.
\end{equation}
Here $v = c/\sqrt{\epsilon_r}$ and the factor $-i \omega^{-1}$ has been chosen so that $\psi_{\nu,\omega}$ here corresponds to $\psi_{\nu,\omega}$ in equation (2) in the main text. This leads to an equation of motion which is first-order in $z$ (assuming the envelope varies slowly),
\begin{equation}\label{Schrodinger equation}
i \partial_z \psi_{\nu,\omega}(x,y,z) = -\frac{1}{2} \bigg( \frac{\omega}{v} \bigg)^{-1} \nabla^2_\perp \psi_{\nu,\omega}(x,y,z) -  \frac{1}{2}\bigg( \frac{\omega}{v} \bigg)\, \Delta \epsilon(x,y)\,  \psi_{\nu,\omega}(x,y,z),
\end{equation}
where $\nabla^2_\perp$ is the Laplacian operator acting in the $x$-$y$ plane.  This resembles a two-dimensional Schr\"odinger equation with the $z$-direction taking the place of time, and with a potential $\Delta \epsilon(x,y)$. Motivated by this analogy, we translate the Schr\"odinger-like equation into a lattice model by writing $\psi(x,y,z)$ (we will suppress the $\nu$ and $\omega$ indices for now) as a sum over ``Wannier states" of the lattice: $\psi(x,y,z) = \sum_{\bm{r}} \psi_{\bm{r}}(z) w_{\bm{r}} (x,y)$. Using the orthonormality of Wannier states we find
\begin{equation}
i \partial_z \psi_{\bm{r}}(z) = \psi_{\bm{r}}(z)\, \langle w_{\bm{r}}|H|w_{\bm{r}}\rangle + \sum_{\bm{r}' \neq \bm{r}} \psi_{\bm{r}'}(z)\,  \langle w_{\bm{r}}|H|w_{\bm{r}^\prime}\rangle,
\end{equation}
where we have switched to bra-ket notation for the Wannier states, and $H = -\frac{1}{2} ( \frac{\omega}{v} )^{-1} \nabla_\perp^2  -  \frac{1}{2}( \frac{\omega}{v} ) \Delta \epsilon(x,y) $. Matrix elements between different Wannier states decay exponentially with the spacing between lattice sites, and so we neglect all off-diagonal elements except $H_{\bm{r},\bm{r}+\bm{s_{j}}} = \langle w_{\bm{r}}|H|w_{\bm{r}+\bm{s_{j}}}\rangle$. We remove the diagonal elements, which adjust the wavevector to satisfy the massive dispersion relation of the waveguide at the ``site" $\bm r$, by defining $k_{\omega,\bm r} \equiv \omega/v + \langle w_{\bm{r}}|H|w_{\bm{r}}\rangle$ and replacing $\omega/v \rightarrow k_{\omega}$ in equation~(\ref{Atopsi}). With these adjustments, we arrive at the tight-binding equation for $\psi_{\bm{r}}(z)$,
\begin{equation}\label{tight binding 1}
i \partial_z \psi_{\bm{r}}(z)  = \sum_{\bm{j} = 1,2,3}  H_{\bm{r},\bm{r}\pm\bm{s_{j}}} \psi_{\bm{r}\pm\bm{s_{j}}}(z).
\end{equation}
Here the sign choice is positive if $\bm{r}$ is on sublattice A and negative if on sublattice B. Restoring indices, and now using $\nu$ to index eigenvectors of this Hamiltonian, we have
\begin{equation}
\psi_{\nu,\omega}(x,y,z) = \sum_{\bm{r}} e^{-i \kappa_{\nu,\omega} z}\, \psi_{\nu,\omega,\bm{r}}\, w_{\bm{r}} (x,y).
\end{equation}
where $\psi_{\nu,\omega,\bm{r}}$ is an eigenstate of the Hamiltonian $H_{\omega,\bm{r},\bm{r}\pm\bm{s_{j}}}$ with eigenvalue $\kappa_{\nu,\omega}$. Our full expressions for the vector potential and electric field are
\begin{equation}
\hat{\bm{A}} (\bm{r},t)  = \sum_{ \nu,\omega} (-i \omega^{-1})\sum_{\bm{r}} e^{i(k_{\omega,\bm r} - \kappa_{\nu,\omega})z - i\omega t}\, \psi_{\nu,\omega,\bm{r}}\, w_{\bm{r}}(x,y)\,  \hat{b}_{\nu,\omega}\, \bm{n} + \text{H.c.},
\end{equation}
\begin{equation}
\hat{\bm{E}} (\bm{r},t) = -\frac{1}{\epsilon(\bm{r})\epsilon_0} \hat{\bm{\Pi}}  = \sum_{ \nu,\omega} \sum_{\bm{r}} e^{i(k_{\omega,\bm r} - \kappa_{\nu,\omega})z - i\omega t}\, \psi_{\nu,\omega,\bm{r}}\, w_{\bm{r}}(x,y)\,  \hat{b}_{\nu,\omega}\, \bm{n} + \text{H.c.}
\end{equation}
With these solutions in hand, we can derive commutation relations of the $\hat{b}_{\nu,\omega}$ and $\hat{b}^{\dagger}_{\nu,\omega}$ by enforcing our commutation relations for $\hat{\bm{A}}$ and $\hat{\bm{\Pi}}$. As expected, we find that:
\begin{equation}
[\hat{b}_{\nu,\omega},\hat{b}_{\nu^\prime,\omega^\prime}] = 0\\
\end{equation}
\begin{equation}
[\hat{b}^{\dagger}_{\nu,\omega},\hat{b}^{\dagger}_{\nu^\prime,\omega^\prime}] = 0\\
\end{equation}
\begin{equation}
[\hat{b}_{\nu,\omega},\hat{b}^{\dagger}_{\nu^\prime,\omega^\prime}] = \delta_{\nu,\nu'} \delta_{\omega,\omega'}.
\end{equation}
Lastly, it is convenient to decompose $\hat{\bm{E}} (\bm{r},t)$ in the Wannier basis, and define an electric field operator at each lattice site using:
\begin{equation}
\hat{\bm{E}} (\bm{r},t) =  \sum_{\bm{r}} \hat{\bm{E}}_{\bm{r}}(z,t)\,  w_{\bm{r}}(x,y) ,
\end{equation}
\begin{equation}
\hat{\bm{E}}_{\bm{r}}(z,t) =  \sum_{ \nu,\omega} e^{i(k_{\omega,\bm r} - \kappa_{\nu,\omega})z - i\omega t}\, \psi_{\nu,\omega,\bm{r}}\, \hat{b}_{\nu,\omega} \, \bm{n} + \text{H.c.}
\end{equation}
In the main text, we have specialized to the case of monochromatic input light, which selects a single frequency $\omega$ in the sum above.  

For the sake of simplicity, we have neglected in the main text the $\bm r$-dependence of the waveguide dispersion $k_{\omega,\bm r}$.  This is rigorously justifiable when the variation in the relative permittivity $\Delta\epsilon(x,y)$ has the discrete translational symmetry of the undistorted hexagonal lattice, or that of the translationally-invariant Kekul\'e pattern shown in Fig.~1(A).  However, when translational symmetry is broken by the presence of a vortex, $k_{\omega,\bm r}$ is not independent of the waveguide position $\bm r$.  This is problematic for systems with multiple vortices, as it can lead to the accumulation of dynamical phases as the vortices are braided (see Appendix E).  We argue in Appendix E that the effects of these dynamical phases on the zero modes can be eliminated for an appropriate class of distortions of the waveguide lattice, presented in Appendix C.

\section{Zero-mode statistics and exact diagonalization results}

\subsection{Hamiltonian}
We begin by reviewing the geometry of the hexagonal waveguide lattice.  The vectors connecting a site on sublattice $A$ with its three nearest-neighbors are 
\begin{equation}\label{NN vecs}
\bm{s}_1 = \big(0,\ -a\big),\ \bm{s}_2 = \big(\frac{\sqrt{3}}{2}\,a,\ \frac{1}{2}\,a\big),\text{ and }\bm{s}_3 = \big(\frac{-\sqrt{3}}{2}\, a,\ \frac{1}{2}\, a\big).
\end{equation} 
A unit cell consists of one site on sublattice $A$ and a single nearest-neighbor on sublattice $B$. Our lattice vectors are then 
\begin{equation}\label{lattice vecs}
\bm{a}_1 = \bm{s}_2 - \bm{s}_1 = \big(\frac{\sqrt{3}}{2}\, a,\ \frac{3}{2}\, a\big),\ \bm{a}_2 = \bm{s}_2 - \bm{s}_3 =\big(\sqrt{3}\, a,\ 0\big).
\end{equation}
 We have chosen $a$ to be the distance between nearest-neighbors, whereupon the distance between unit cells is $\sqrt{3}\, a$. In reciprocal space, the Dirac points lie at opposite corners of the hexagonal Brillouin zone, $\bm{K}_{\pm} = \pm \big(\frac{4\pi}{3\sqrt{3}a},\, 0\big)$.

The Hamiltonian for the class of waveguide lattices that we study is given by
\begin{equation}\label{tight binding}
H_{\bm{r},\bm{r}+\bm{s_{j}}}  = -t - \delta t_{\bm{r},i},
\end{equation}
\begin{equation}
H_{\bm{r}+\bm{s_{j}},\bm{r}}  = H_{\bm{r},\bm{r}+\bm{s_{j}}}
\end{equation}
for $\bm{r}$ in sublattice $A$ and all other matrix elements zero. In the presence of a Kekul\'{e} texture, $\delta t_{\bm{r},i}$ is given in terms of the complex-valued order parameter $\Delta(\bm{r})$ by:
\begin{equation}\label{delta_t}
\delta t_{\bm{r},i} = \frac{1}{2}\, \Delta(\bm{r})\, e^{i\bm{K}_{+} \cdot \bm{s}_i}\,e^{i(\bm{K}_{+}-\bm{K}_{-}) \cdot \bm{r}} + \text{c.c.}
\end{equation}
A vortex centered at position $\bm{R}$ is described by $\Delta(\bm{r};\bm{R}) = \Delta_0(|\bm{r}-\bm{R}|)e^{i [ \alpha - \text{arg}(\bm{r}-\bm{R})]}$, where $\Delta_0(|\bm{r}-\bm{R}|)$ is a real-valued function which vanishes at $|\bm{r}-\bm{R}| = 0$ and approaches a constant value when $|\bm{r}-\bm{R}| \gg \ell_0$, for some length scale $\ell_0$.  For simplicity, we take $\Delta_0(|\bm{r}-\bm{R}|)=\Delta\, \tanh(|\bm r - \bm R|/\ell_0)$, as in the main text. For multiple vortices, the order parameter is proportional to the product of order parameters for each individual vortex; for $v$ identical vortices at positions $\bm{R}_1,\dots,\bm{R}_v$, we have

\begin{equation}\label{n vort texture}
\Delta(\bm{r}) = \Delta\bigg[ \prod_{j=1}^{v} \tanh(|\bm{r}-\bm{R}_j|/\ell_0) \bigg] e^{i \sum_{j=1}^{v} [\alpha_j - \text{arg}(\bm{r}-\bm{R}_j)]}.
\end{equation}

\subsection{Vortex statistics}

The Kekul\'{e} texture with $v$ vortices admits $v$ zero modes, one localized at each vortex. We assume vortices are kept far enough apart to prevent mixing between zero modes (see Appendix E). Near a single vortex, say near $\bm{R}_1$, we can approximate $\bm{r} \approx \bm{R}_1$ in equation~(\ref{n vort texture}) for all contributions from vortices $\bm{R}_j,j>1$. So the zero mode at the first vortex sees an order parameter
\begin{equation}
\Delta(\bm{r}) \approx \Delta\tanh(|\bm{r}-\bm{R}_1|/\ell_0)\bigg[ \prod_{j=2}^{v} \tanh(|\bm{R}_1-\bm{R}_j|/\ell_0) \bigg] e^{ -i\, \text{arg}(\bm{r}-\bm{R}_1) } e^{ i\alpha_1 + i\sum_{j=2}^{v} [ \alpha_j - \text{arg}(\bm{R}_1-\bm{R}_j)]}\nonumber
\end{equation}
\begin{equation}\label{local approx}
\hspace{-1.15in}\approx \Delta\tanh(|\bm{r}-\bm{R}_1|/\ell_0)\, e^{ -i\, \text{arg}(\bm{r}-\bm{R}_1) } e^{ i\alpha_1 + i\sum_{j=2}^{v} [ \alpha_j - \text{arg}(\bm{R}_1-\bm{R}_j)]},
\end{equation}
where we have used that, for $\ell_0$ much smaller than the distance between vortices, we have $\tanh(|\bm{R}_i-\bm{R}_j|/\ell_0)\approx1$ for all $i,j$. Thus, the only effect of the $n-1$ other vortices on the first zero mode is to shift the phase of its order parameter by a constant, \textit{i.e.}~taking $\alpha_1 \rightarrow \alpha_1 + \sum_{j=2}^{v} [ \alpha_j - \text{arg}(\bm{R}_1-\bm{R}_j)]$.

Now consider braiding two vortices at positions $\bm{R}_1$ and $\bm{R}_2$ in a counterclockwise fashion. The braiding operation takes $\text{arg}(\bm{R}_1-\bm{R}_2) \rightarrow \text{arg}(\bm{R}_1-\bm{R}_2) + \pi$ and $\text{arg}(\bm{R}_2-\bm{R}_1) \rightarrow \text{arg}(\bm{R}_2-\bm{R}_1) + \pi$, while leaving all other $\text{arg}(\bm{R}_i-\bm{R}_j)$ unchanged. Using the local approximation of the order parameter near $\bm{R}_1$ in equation~(\ref{local approx}), we see that braiding vortices $\bm{R}_1$ and $\bm{R}_2$ is equivalent to taking $\alpha_1 \rightarrow \alpha_1 - \pi$ in the single vortex wavefunction. Using a similar approximation near the vortex at $\bm{R}_2$, we also have $\alpha_2 \rightarrow \alpha_2 - \pi$. Relabeling the exchanged vortices $1 \leftrightarrow 2$ and using $\text{arg}(\bm{R}_1-\bm{R}_2) = \text{arg}(\bm{R}_2-\bm{R}_1) + \pi$, we finally have $\alpha_1 \rightarrow \alpha_2$ and $\alpha_2 \rightarrow \alpha_1 - 2\pi$. Because the zero mode wavefunction (4) is double-valued with respect to $\alpha$, we have $u(r; \alpha - 2\pi) = -u(r;\alpha)$, and we arrive at the statistics in Eq.~(8).

\subsection{Tight-binding braiding simulation}
We here review exact diagonalization calculations of the tight-binding Hamiltonian~(\ref{tight binding})-(\ref{n vort texture}), to verify the existence of the zero modes and the action of the braiding generators in Eq. (8). Upon introduction of a constant $\Delta(\bm{r}) = \Delta$, a band gap opens among the bulk states, although a small number of edge states with zero and near-zero energy remain. The introduction of a vortex in the order parameter leads to a single zero mode localized at the vortex core, and an additional edge mode, which are decoupled from one another and at zero energy for sufficiently large lattices. The zero mode has support only on sublattice $A$, and decays exponentially with distance from the vortex core with a characteristic length inversely related to $|\Delta|$.

\begin{figure}
\centering
a)\, \includegraphics[width=.4\textwidth,page=5]{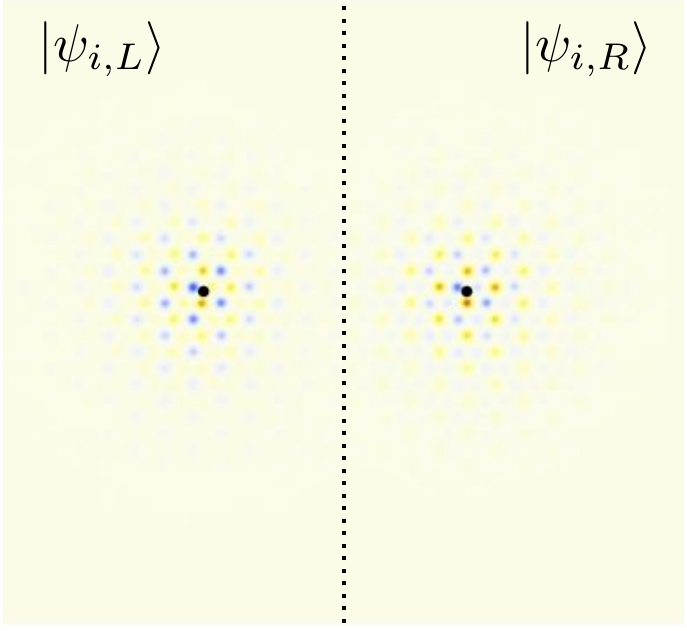}
\hspace{1mm} 
b)\, \includegraphics[width=.4\textwidth,page=6]{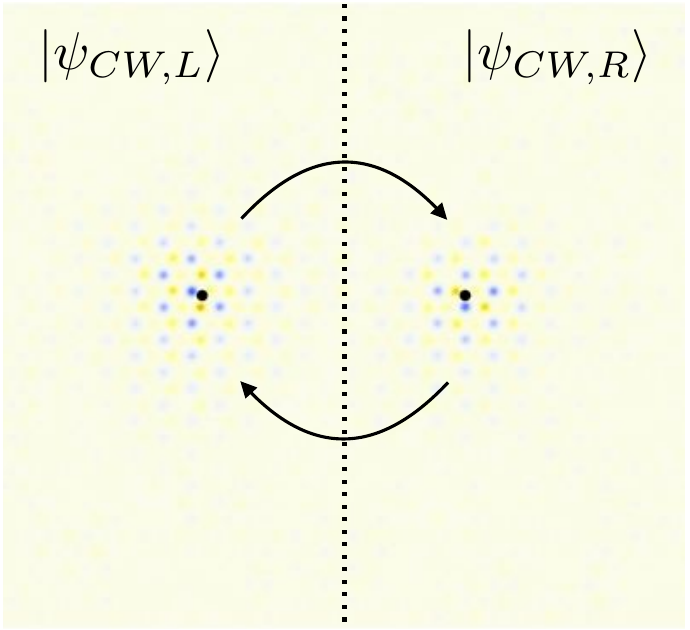}\\
\vspace{5mm}
c)\, \includegraphics[width=.4\textwidth,page=7]{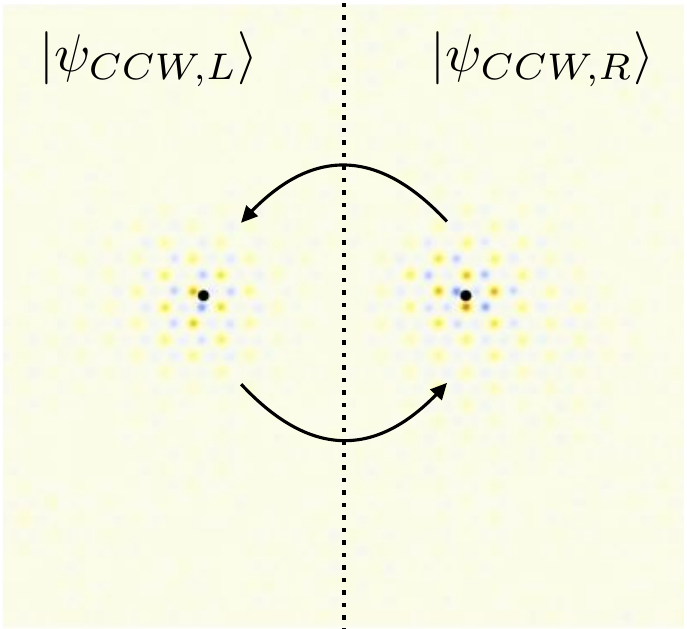}\ \includegraphics[width=.069\textwidth,page=8]{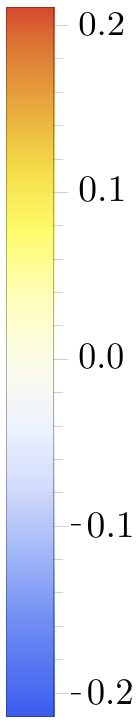}
\caption{(a) The superposition of vortex zero modes used as the initial state for both braids. Black dots are placed at the vortex centers, and a dashed line is drawn at the boundary used to decompose the initial and final wavefunctions into left and right pieces, as discussed in the text. To display the results in a more visually-friendly manner, the tight-binding wavefunction amplitude was multiplied by an arbitrarily chosen Gaussian ``Wannier function'' at each lattice site. (b) Results of the tight-binding simulation of a single CW braid. Arrows indicate the handedness of the braid performed. A visual comparison to (a) shows that the braid has resulted in a left vortex of the same sign as the initial state, and a right vortex of opposite sign, confirming the action of the braiding operators (8). (c) Results of the tight-binding simulation of a single CCW braid. The left vortex is now of opposite sign as the initial state, and the right vortex is of the same sign.} \label{fig: tbresults}
\end{figure}

Again exploiting the mapping between the Helmholtz equation~(\ref{helmholtz}) and the tight-binding equation~(\ref{tight binding 1}), we simulate the effect of braiding in our waveguide array by promoting our tight-binding Hamiltonian ~(\ref{tight binding}) to a function of time, $H(t)$ (recall that ``time" in the tight-binding equation is actually the spatial $z$-direction in the waveguide array --- we refer to it as $t$ in this section to make it clear that our simulations were performed in a two-dimensional system), implicitly through its dependence on $\Delta(\bm{R})$ and thus the vortex centers $\bm{R}_i(t)$. We simulate a system where $\Delta(\bm{r})$ hosts two vortices, at positions $\bm{R}_1$ and $\bm{R}_2$. To perform a single counterclockwise (CCW) or clockwise (CW) braid of the vortices, we describe the vortex positions as a function of time by
\begin{equation}
\bm{R}_1(t) = R_B \, \big( \cos(\pi t/T), \, \pm \sin(\pi t/T) \big), \\
\end{equation}
\begin{equation}
\bm{R}_2(t) = -R_B \, \big( \cos(\pi t/T), \, \pm \sin(\pi t/T) \big), \\
\end{equation}
where $R_B$ is the radius of the semicircle traced out by each vortex during the braid (or equivalently, half the distance between the vortices), $T$ is the total time of the process, and the sign of the $y$-component is determined by the handedness of the braid. The initial state of the system was taken to be an equal-amplitude superposition of the zero mode states centered at each of the vortices. We then discretize the time interval into $N$ steps of duration $\delta t = T/N$, and act with the matrix $e^{iH(t_j) \delta t}$ on the state at each time $t_j=j\, \delta t$, $j=1,\dots,N$, achieving the final state after $t_j = N\, \delta t = T$.

Results for a simulation on a $44\times45$-unit-cell lattice with $R_B=12\, a$, $T=120/t$, $N = 100$, and $\Delta = .7\, t$ are shown in Fig.~(\ref{fig: tbresults}). The braiding generators (8) predict that under a CW braid, the vortex on the right undergoes a sign change between the initial and final states, while the vortex on the left does not.  For a CCW braid, on the other hand, the vortex on the left undergoes a sign change, while the vortex on the right does not.  A visual comparison of the final wavefunction after each braid to the initial wavefunction confirms this action (see Fig.~\ref{fig: tbresults}). To quantify this, it is helpful to decompose each wavefunction as $|\psi \rangle = |\psi_L \rangle + |\psi_R \rangle$, where $|\psi_L \rangle$ has support only to the left of a vertical line equidistant from both vortices, and $|\psi_R \rangle$ has support only to the right. We write the initial state of both braids as $|\psi_i \rangle$ and the final state after CW/CCW braids as $|\psi_{CW} \rangle$ and $|\psi_{CCW} \rangle$, respectively. In the ideal limit of well-separated vortices and adiabatic braiding, the braiding generators predict that $\langle \psi_{i,L} |\psi_{CW,L} \rangle = \langle \psi_{i,R} |\psi_{CCW,R} \rangle = 0.5$, and $\langle \psi_{i,R} |\psi_{CW,R} \rangle = \langle \psi_{i,L} |\psi_{CCW,L} \rangle = -0.5$. Note that the overlaps have magnitude less than unity because the $|\psi_L \rangle$ and $|\psi_R \rangle$ are not normalized states. Computing the overlaps numerically, we find
\begin{subequations}
\begin{align}
\langle \psi_{i,L} |\psi_{CW,L} \rangle &= 0.47, \label{signs a}\\
\langle \psi_{i,R} |\psi_{CW,R} \rangle &= -0.39, \label{signs b}\\
\langle \psi_{i,L} |\psi_{CCW,L} \rangle &= -0.38, \label{signs c}\\
\langle \psi_{i,R} |\psi_{CCW,R} \rangle &= 0.46, \label{signs d}
\end{align}
\end{subequations}
with zero imaginary part (to one part in $10^{-12}$--$10^{-14}$) for all overlaps.  Note that, as predicted, the vortex on the left does not undergo a sign change under a CW exchange, while the vortex on the right does [c.f.~Eqs.~\eqref{signs a} and \eqref{signs b}].  Similarly, as predicted, the vortex on the left acquires a sign change under a CCW exchange, while the vortex on the right does not [c.f.~Eqs.~\eqref{signs c} and \eqref{signs d}].  Thus, the signs of the numerically-obtained overlaps \eqref{signs a}--\eqref{signs d} are consistent with the analytic predictions obtained from the action of the braiding generators (8).  However, note that the magnitudes of the overlaps are not 0.5, as would be expected for perfectly adiabatic braids.

We identify two reasons for these deviations from ideal behavior. The first is non-adiabatic loss to bulk modes of the lattice, which decreases the magnitude of all overlaps between the initial and final states and results from a failure to satisfy adiabaticity (see Appendix E.2 for a detailed discussion of the adiabatic condition). The second source of non-ideal behavior is adiabatic loss \textit{between} vortices, which accounts for the differing magnitudes between left and right overlaps of the same braid. The two vortex zero modes form a degeneracy-2 subspace of the total Hilbert space of the system, and in principle nonzero Berry's matrix elements between these modes (or equivalently, between the symmetric and antisymmetric superpositions of these modes) are allowed. Such matrix elements would lead to a final state with left and right decompositions of different magnitude, consistent with the following numerically-computed norms:
\begin{subequations}
\begin{align}
\langle \psi_{CW,L} |\psi_{CW,L} \rangle = 0.56, \label{overlaps a}\\
\langle \psi_{CW,R} |\psi_{CW,R} \rangle = 0.44, \label{overlaps b}\\
\langle \psi_{CCW,L} |\psi_{CCW,L} \rangle = 0.42, \label{overlaps c}\\
\langle \psi_{CCW,R} |\psi_{CCW,R} \rangle = 0.58. \label{overlaps d}
\end{align}
\end{subequations}
[Note that Eqs.~\eqref{overlaps a} and \eqref{overlaps b} add up to 1, as do Eqs.~\eqref{overlaps c} and \eqref{overlaps d}, as expected.]  We expect the Berry's matrix elements leading to the transfer of intensity between vortices to vanish when the vortices are sufficiently far apart. However, this requires the vortices to traverse longer distances, which hinders the adherence to the adiabatic condition. In principle this could be compensated by increasing the time $T$ of the braiding process or the system size; in practice, this leads to a substantial increase in the runtime of the simulation. Note that neither of these sources of non-ideal behavior affects the phase of the computed overlaps, which remain consistent with the overlaps predicted by the braiding generators.

%

\subsection{Derivation of the braiding generators in Eq.~(8)}

In this section, we show that the operators
\begin{equation}
  \hat{\tau}_i=e^{\pi
    \left(
    \hat{b}_{0,i+1}^\dagger\,\hat{b}_{0,i}-
    \hat{b}_{0,i}^\dagger\,\hat{b}_{0,i+1}
    \right)/2}
  \;,
  \label{tau_def}
\end{equation}
defined in Eq.~(8) of the main text, implement the braiding operations
\begin{subequations}\label{appendix braiding action}
\begin{eqnarray}
  \hat{\tau}_i\;\hat{b}_{0,i}\;\hat{\tau}_i^\dagger&=&\hat{b}_{0,i+1} \label{tau_A}\\
  \hat{\tau}_i\;\hat{b}_{0,i+1}\;\hat{\tau}_i^\dagger&=&-\hat{b}_{0,i} \label{tau_B}\\
  \hat{\tau}_i\;\hat{b}_{0,j}\;\hat{\tau}_i^\dagger&=&\hat{b}_{0,j}\;,\quad j\ne i,i+1\;\label{tau_C}
\end{eqnarray}
\end{subequations}
explored above.  The third line (\ref{tau_C}) follows because if $j\ne i$ or $i+1$, then
$[\hat{b}_{0,j},\hat{\tau}_i]=0$, since $\hat{\tau}_i$ in (\ref{tau_def}) only contains bosons
with labels $i$ and $i+1$.  Note that one can use Eqs.~\eqref{tau_def} and \eqref{appendix braiding action} to show that the braiding generators $\hat{\tau}_i$ and $\hat{\tau}_j$ do not commute for all $i$ and $j$, thus demonstrating directly that the $\hat{\tau}_i$ form a non-Abelian representation of the braid group.

To show lines (\ref{tau_A}) and (\ref{tau_B}), it is convenient to make the change of variables
\begin{equation}
  \hat{\beta}_{i,\pm}=\frac{1}{\sqrt{2}}\left(\hat{b}_{0,i}\pm i\,\hat{b}_{0,i+1}\right)
  \;,
\end{equation}
writing
\begin{equation}
  \hat{b}_{0,i}=\frac{1}{\sqrt{2}} \left(\hat{\beta}_{i,+}+\hat{\beta}_{i,-}\right)
  \quad
  \hat{b}_{0,i+1}=\frac{1}{\sqrt{2}\,i} \left(\hat{\beta}_{i,+}-\hat{\beta}_{i,-}\right)
  \;.
  \label{b-in-terms-of-beta}
\end{equation}
Notice that, as defined, $[\hat{\beta}_{i,\pm},\hat{\beta}_{i,\pm}^\dagger]=1$, and
$[\hat{\beta}_{i,\pm},\hat{\beta}_{i,\mp}^\dagger]=0$. In terms of the new bosonic operators $\hat{\beta}_{i,\pm}$, we can write
\begin{equation}
  \hat{\tau}_i=e^{i\frac{\pi}{2}\,
    \left(
    \hat{\beta}_{i,+}^\dagger\,\hat{\beta}_{i,+}-
    \hat{\beta}_{i,-}^\dagger\,\hat{\beta}_{i,-}
    \right)}
  \;.
\end{equation}
Now,
$
e^{i\frac{\pi}{2}\left(
  \hat{\beta}_{i,+}^\dagger\,\hat{\beta}_{i,+}
  \right)}\,\hat{\beta}_{i,+}
=
e^{-i\frac{\pi}{2}}
\;
\hat{\beta}_{i,+}\,e^{i\frac{\pi}{2}\left(
  \hat{\beta}_{i,+}^\dagger\,\hat{\beta}_{i,+}
  \right)}
$
and
$
e^{-i\frac{\pi}{2}\left(
  \hat{\beta}_{i,-}^\dagger\,\hat{\beta}_{i,-}
  \right)}\,\hat{\beta}_{i,-}
=
e^{+i\frac{\pi}{2}}
\;
\hat{\beta}_{i,-}\,e^{-i\frac{\pi}{2}\left(
  \hat{\beta}_{i,-}^\dagger\,\hat{\beta}_{i,-}
  \right)}
$,
so that
\begin{subequations}
  \begin{eqnarray}
  \hat{\tau}_i\;\hat{\beta}_{i,+}\;\hat{\tau}_i^\dagger&=& -i\hat{\beta}_{i,+} \\
  \hat{\tau}_i\;\hat{\beta}_{i,-}\;\hat{\tau}_i^\dagger&=&+i\hat{\beta}_{i,-} \;,
\end{eqnarray}
\end{subequations}
from which (\ref{tau_A}) and (\ref{tau_B}) follow using
(\ref{b-in-terms-of-beta}).

\section{Waveguide worldlines for vortex braiding}

In this section, we provide explicit formulas for the $x$-$y$ positions of the waveguides in the presence of $v$ vortices in the Kekul\'e order parameter.  The positions of the centers of the waveguides in the hexagonal lattice are given by
\begin{equation}
\bm r_A = m_1\, \bm a_1+m_2\, \bm a_2,\qquad \bm r_B = \bm r_A+\bm s_1,
\end{equation}
where $m_{1,2}$ are integers, the vectors $\bm a_{1,2}$ spanning the triangular lattice are given in Eq.~(\ref{lattice vecs}), and the vector $\bm s_1$ connecting sublattices A and B is defined in Eq.~(\ref{NN vecs}).

A Kekul\'e distortion, constant or spatially varying, displaces the equilibrium positions of the waveguides to new positions $\bm r_{A,B}+\bm u_{A,B}(\bm r_{A,B})$, with
\begin{equation}
\bm u_A(\bm r)=\frac{i}{2}\, \gamma\,\Delta(\bm r)\, e^{-i\bm{r}\cdot\bm K_+}\left(\begin{matrix}1 \\+ i\end{matrix}\right) + \text{c.c.}
\end{equation}
\begin{equation}
\bm u_B(\bm r)=\frac{i}{2}\, \gamma\,\Delta(\bm r)\, e^{-i\bm{r}\cdot\bm K_+}\left(\begin{matrix}1 \\ -i\end{matrix}\right) + \text{c.c.},
\end{equation}
where $\gamma$ is a dimensionful constant with units of $[\text{length}]^2$. The pattern of waveguide displacements in the presence of $v$ vortices is obtained by substituting Eq.~(\ref{n vort texture}) into the above.

In this work, we have modified slightly the above pattern of distortions in the presence of $v$ vortices, in order to mitigate the effect of the position-dependence of the waveguide dispersion $k_{\omega,\bm r}$ on the braiding of the associated zero modes.  The modified pattern of distortions used to generate Fig. 1(B), which leads to a hopping modulation of the form (\ref{delta_t}) with an order parameter of the form (\ref{n vort texture}), is
\begin{equation}\label{modified displacements}
\bm u_A(\bm r)=\frac{i}{2}\, \gamma\,\Delta(\bm r)\, e^{-i\bm{r}\cdot\bm K_+}\left(\begin{matrix}1 \\+ i\end{matrix}\right) + \text{c.c.},\qquad \bm u_B(\bm r)=0.
\end{equation}
We show in Appendix D that, for small displacements $\bm u_{A,B}$, this choice of lattice distortion leads to an onsite potential that is uniform for all sites in sublattice $A$, so that the localized states in the vortex core, which have support only in sublattice $A$, experience an onsite potential that is uniform everywhere.  (For a vortex with the opposite vorticity, whose associated zero mode has support on sublattice $B$ rather than $A$, one would need to freeze sublattice $A$ instead of sublattice $B$.)

To write into some bulk medium a photonic lattice that implements an arbitrary braid $\hat B$, one simply promotes the displacements $\bm u_{A,B}(\bm r)\to\bm u_{A,B}(\bm r,z)$ by introducing a $z$-dependence of the vortex positions $\bm R_i\equiv \bm R_i(z)$ in Eq.~(\ref{n vort texture}).  The $\bm R_{i}(z)$ can be defined piecewise on intervals of length $L_{B}$ (see Appendix E) for each segment of the braid corresponding to the application of a single braiding generator.  On each of these intervals, $\bm R_i(z)$ can be chosen to be an arbitrary parametric function of $z$ that interpolates between the initial and final positions of the $i$-th vortex during that segment of the braid (albeit sufficiently slowly that the adiabatic condition holds).  Examples of such piecewise parametric functions are shown in Fig. 2 of the main text, where the strands being braided represent the vortex-core positions $\bm R_{i}(z)$ for $i=1,2,3$.

\section{Comparison with Majorana zero modes}

\begin{figure}
\centering
a)\, \includegraphics[width=.2\textwidth,page=9]{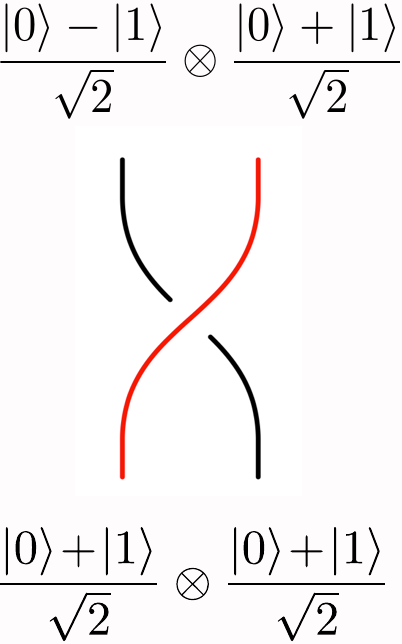}
\hspace{3cm} 
b)\, \includegraphics[width=.2\textwidth,page=10]{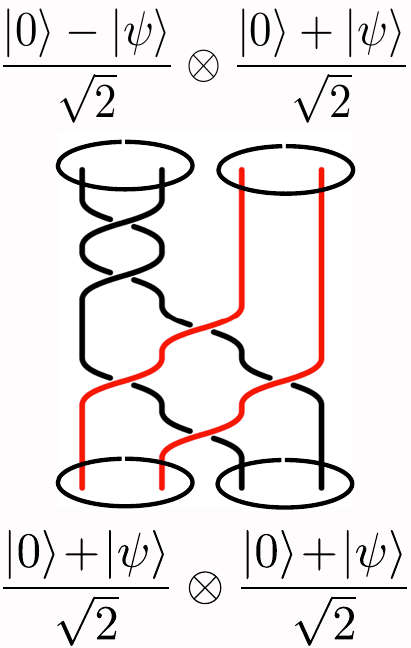}
\caption{Simulating the braiding of photonic zero modes with Majorana zero modes.  To reproduce the result of a clockwise exchange of two photonic zero modes [panel (a)], it is necessary to perform a clockwise exchange of two \textit{pairs} of Majorana zero modes, followed by a full $2\pi$-braid of one of the pairs [panel (b)].  The two processes coincide upon identifying the states $\ket 0$ and $\ket 1$ in panel (a) with the states $\ket 0$ and $\ket \psi$ in panel (b). \label{fig: simulation with majoranas}}
\end{figure}

The action \eqref{appendix braiding action} of the braiding generators \eqref{tau_def} on the zero-mode annihilation operators $\hat{b}_{0,i}$ is identical to the action of the corresponding braiding generators for Majorana zero modes (c.f., e.g., Ref.~\cite{Ivanov}).  The question then naturally arises: how similar to Majorana zero modes are the photonic zero modes studied in this work?  It turns out, as we show below, that the braiding of these photonic zero modes can be simulated by braiding \textit{pairs} of Majorana zero modes. The converse of this statement is not true --- photonic zero modes cannot simulate Majorana zero modes. Nevertheless, their braiding is still non-Abelian, as evidenced by the noncommutativity of the braiding generators \eqref{tau_def}.

To understand how to simulate the braiding of photonic zero modes with Majoranas, first recall that there is a well-defined notion of parity for the photonic zero modes: a single zero mode can be occupied by either an even or an odd number of photons, and this even-odd distinction affects the outcomes of braiding processes [c.f.~the discussion following Eq.~(10) in the main text].  For Majorana zero modes, on the other hand, only \textit{pairs} of zero modes have a definite fermion parity.  This suggests that it should be possible to emulate the braiding of photonic zero modes with pairs of Majoranas.

To see that this is indeed possible, consider performing the braid described by Eq.~(11) in the main text--namely, the clockwise exchange of two photonic zero modes, each of which is initially in the state $(\ket{0}+\ket{1})/\sqrt 2$, as shown in Fig.~\ref{fig: simulation with majoranas}(a).  (Here, as before, $\ket{0}$ and $\ket{1}$ are single-vortex states with zero and one photon occupying the zero mode.  One could obtain the same result by replacing the state $\ket 0$ with a state $\ket{\rm even}$ with an even number of photons, and the state $\ket{1}$ with a state $\ket{\rm odd}$ with an odd number of photons.)  To perform the same operation with Majoranas, one must start with two pairs of Majoranas, each of which is initially in the state $(\ket{0}+\ket{\psi})/\sqrt 2$, where $\ket{0}$ is the two-Majorana state with even fermion parity and $\ket{\psi}$ is the two-Majorana state with odd fermion parity.  Then, to simulate the exchange of two photonic vortices, one must exchange the two pairs of Majoranas, and then perform a full $2\pi$-braid of one of the pairs, as shown in Fig.~\ref{fig: simulation with majoranas}(b).  As indicated in the figure, the initial and final states of the two processes are identical upon identifying the states $\ket 0$ and $\ket 1$ in the photonic system with the states $\ket 0$ and $\ket \psi$ in the Majorana system.  In other words, we have shown that the set of operations that can be performed by braiding photonic zero modes is a \textit{subset} of the set of operations that can be performed by braiding Majorana zero modes.  This subset is defined by imposing a constraint whereby Majoranas can only be braided in pairs, with each braid being of the type shown in Fig.~\ref{fig: simulation with majoranas}.

This finding has implications for the utility of photonic zero modes in applications for topological quantum computing (TQC).  In particular, since the braiding of Majorana zero modes is not universal for TQC, \cite{NayakReview} it immediately follows that photonic zero modes are also not universal for TQC.  Furthermore, photonic zero modes can perform only a subset of the set of topologically-protected gates that Majoranas can.  Thus, photonic zero modes do not offer an advantage relative to Majorana zero modes for the purposes of TQC.  However, it is a central message of this paper that the control and tunability afforded by photonic platforms presents a marked advantage over solid-state devices in terms of the ability to create, manipulate, and observe topological defects and their non-Abelian braiding.

\section{Justifications for neglecting nonuniversal effects}

In this section we discuss the requirements imposed by dynamical phases, adiabaticity, wavefunction mixing, and finite size effects, and use these to estimate the number of braids we can perform in our system.

\subsection{Dynamical phases}
In our tight-binding Hamiltonian, the sublattice symmetry (SLS) $\psi_{\bm{r}} \rightarrow \psi_{\bm{r}}, \psi_{\bm{r}+\bm{s}_j} \rightarrow -\psi_{\bm{r}+\bm{s}_j}$, with $\bm{r}$ in sublattice $A$, ensures that a zero mode will be at precisely zero energy. However, the presence of a position-dependent waveguide dispersion $k_{\omega,\bm r}$ (referred to below as an onsite potential) or next-nearest-neighbor hopping breaks SLS and can cause the zero mode energy to depend on both $\alpha$ and the position of the vortex center. This could cause the dynamical phase accumulation of the zero mode to depend on the path taken by the vortex, foiling measurement of the geometric phase due to braiding. We show that, and estimate the total dynamical phase accumulated in our system.

\subsubsection{Onsite potential}
We show here that the onsite potential $k_{\omega,\bm r}$ can be made uniform in sublattice $A$ by ``freezing" (\textit{i.e.}, not displacing) the sites on sublattice $B$.  Suppose that $k_{\omega,\bm r}\equiv k_\omega$ in the absence of displacements of the waveguides from the hexagonal lattice sites $\bm r_A$ and $\bm r_B$.  When the waveguides are displaced, one can show that, for $\bm r$ in sublattice $A$,
\begin{align}\label{ultra-local}
k_{\omega,\bm r}=k_\omega\, \exp\left\{-\sum^3_{j=1}\bigg[\Big|\frac{1}{a}\big(\bm s_j-\bm u_A(\bm r)+\bm u_B(\bm r+\bm s_j)\big)\Big|-1\bigg]\right\}.
\end{align}
For $\bm r$ in sublattice $B$, one simply takes $A\to B$ and $\bm s_j$ to $-\bm s_j$ in the above.  Let us now expand this expression to leading order for small displacements, \textit{i.e.}~$|\bm u_{A,B}|\ll a$:
\begin{align}
\frac{k_{\omega,\bm r}}{k_\omega}\approx 1+\frac{1}{a^2}\sum^3_{j=1}\bm s_j\cdot[\bm u_A(\bm r)-\bm u_B(\bm r+\bm s_j)]
\end{align}
Noting that $\sum^3_{j=1}\bm s_j=0$ [\textit{c.f.}~Eq.~(\ref{NN vecs})], we see that $k_{\omega,\bm r}$ depends solely on $\bm u_B$ in this limit.  Therefore, setting $\bm u_B=0$, as in Eq.~(\ref{modified displacements}), renders $k_{\omega,\bm r}\equiv k_\omega$ for $\bm r$ in sublattice $A$, to leading order in the displacements of the waveguides.  (Likewise, setting $\bm u_A=0$ renders $ k_{\omega,\bm r}\equiv k_\omega$ for $\bm r$ in sublattice $B$.)  Keeping this in mind, we proceed to second order in the expansion of $k_{\omega,\bm r}$ for $\bm u_B(\bm r)\equiv 0$, finding
\begin{equation}
\frac{k_{\omega,\bm r}}{k_\omega}\approx 1-\frac{1}{2a^2}\sum^3_{j=1}\left[|\bm u_A(\bm r)|^2-\frac{1}{a^2}\big(\bm s_j\cdot\bm u_A(\bm r)\big)^2\right].
\end{equation}
If $|\bm u_A(\bm r)|$ is constant as a function of $\bm r$, then this leading correction is in fact independent of $\bm r$, \textit{i.e.}~the distortion of the waveguide lattice leads to a position-independent renormalized dispersion $k_{\omega,\bm r}\approx k_{\omega}+\delta k_{\omega}$.  The leading position-dependent correction then appears at third order in $|\bm u_{A}(\bm r)|$.  A position-independent $|\bm u_A(\bm r)|$ can easily be achieved if the vortex width $\ell_0$ is sufficiently small, and if the vortex centers $\bm R_i(z)$ never approach a lattice site to within a distance $\ell_0$ during braiding.

It is also worth pointing out that it is possible to choose the magnitudes $|\bm u_A(\bm r)|$ of the displacements in such a way as to generate a position-independent $k_{\omega,\bm r}$, so long as we take $\bm u_{B}\equiv 0$.  To see this, note that, if $\bm u_B(\bm r)=0$, Eq.~(\ref{ultra-local}) is ``ultra-local" in $\bm r$, as its value at a point $\bm r$ is completely determined by $\bm u_A(\bm r)$.  It is thus sufficient to tune one parameter, namely $|\bm u_A(\bm r)|$, in order to fix $k_{\omega,\bm r}$ to a fixed value for each $\bm r$.

\subsubsection{Next-nearest-neighbor hopping}
\begin{figure}
\centering
\includegraphics[width=.3\textwidth,page=4]{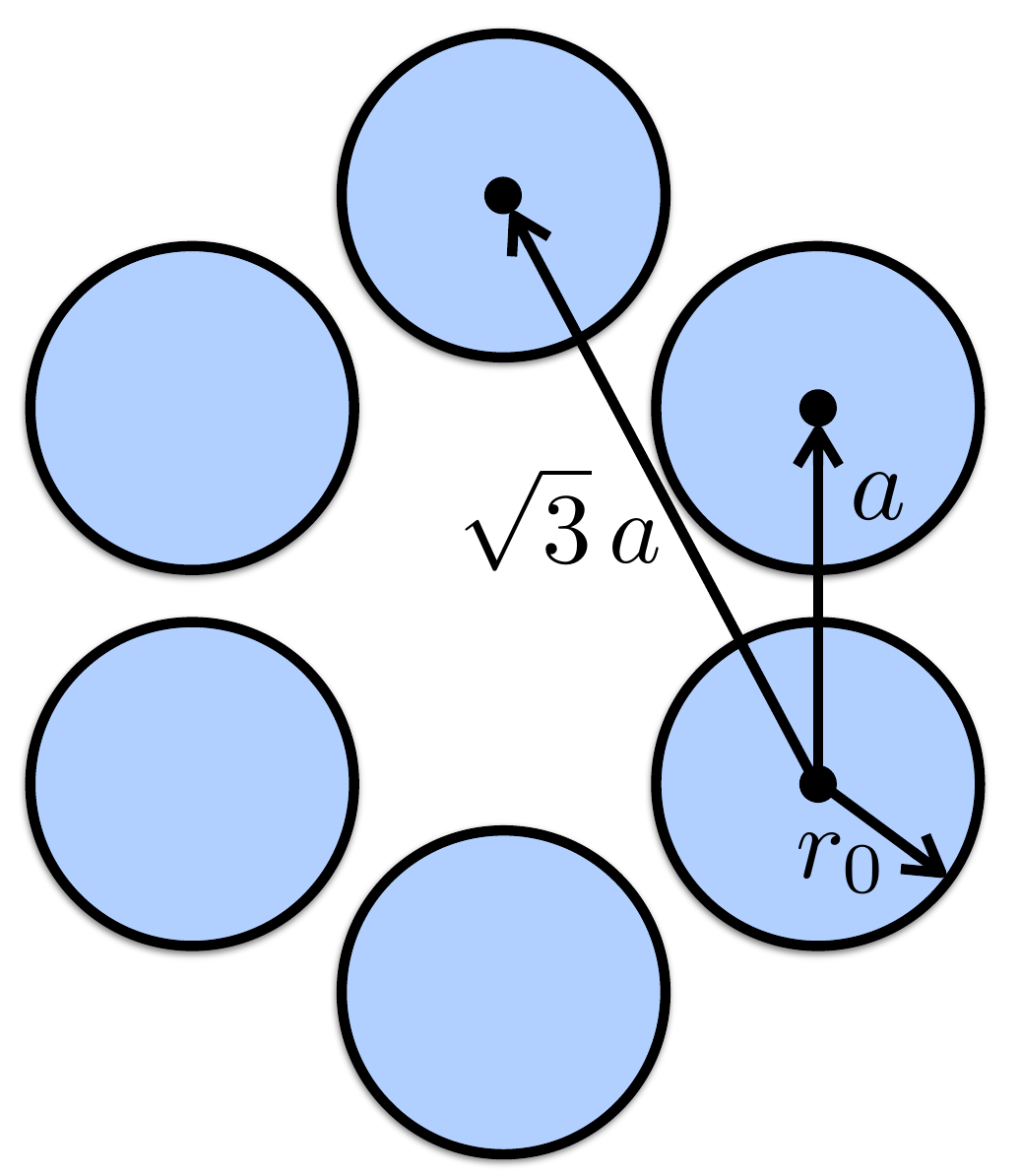}
\caption{Schematic showing the length scales used in Sec.~3$b.$~of this Appendix.  Blue circles represent waveguides. \label{fig: geometry}}
\end{figure}
%
%
%

%

We now outline how the geometry of the waveguide lattice can be tuned such that next-nearest-neighbor hopping $t'$ is strongly suppressed relative to nearest-neighbor hopping $t$. We can estimate $t$ and $t'$ by recalling our Schrodinger equation~(\ref{Schrodinger equation}) for $\psi$. We expect our Wannier states to decay exponentially with decay length $l_0 =\lambda/(2\pi\sqrt{\Delta\epsilon})$, where $\lambda$ is the wavelength of light in the material $\lambda/(2\pi) = v/\omega$. For circular waveguides of radius $r_0$ and nearest-neighbor spacing $a$ (see Fig.~\ref{fig: geometry}), we estimate the hoppings to be

\begin{equation}
t \sim \frac{1}{a} e^{-(a-2r_0)/l_0}
\end{equation}
and
\begin{equation}
t' \sim \frac{1}{a} e^{-(\sqrt{3}a-2r_0)/l_0}.
\end{equation}
The waveguide radius $r_0$ gives us an additional degree of freedom in choosing $t$ and $t'$. We see that if
\begin{equation}
a-2r_0 \approx l_0 \ll (\sqrt{3}-1)a
\end{equation}
holds, one can increase $l_0$ to exponentially suppress the ratio $t'/t$ while only algebraically suppressing $t$.

\subsubsection{Dynamical phase accumulation}

We now estimate the total dynamical phase accumulated during propagation over a vertical distance $L$. As we can exponentially suppress dynamical phases due to next-nearest neighbor hopping, we assume the dominant contribution to the dynamical phase is from a varying onsite potential. Consider a disorder potential that at some waveguide $i$ takes the
form $\mu_i(z)=\eta_i(z)\;\Delta$, where $\eta_i(z)$ is
dimensionless. The energy shift of the zero mode (in the continuum
limit) is
\begin{eqnarray*}
  \delta\epsilon(z)=
  \int d^2x\;\delta\mu(x,z) \;|\psi_0(x)|^2
  \;,
\end{eqnarray*}
and the phase shift after propagation past a distance $L$ is
\begin{eqnarray*}
  \phi= \int_0^L \delta\epsilon(z)\;\frac{dz}{\hbar c}
  \;.
\end{eqnarray*}
The root mean square of the phase shift is
\begin{eqnarray*}
  \delta\phi &=& \eta
  \;\sqrt{\left(\frac{\ell_{xy}}{\xi}\right)^2}
  \;\sqrt{L\,\ell_{z}}
  \;\frac{\Delta}{\hbar c}
  \\
  &\approx& \eta
  \;\frac{\ell_{xy}}{\xi}
  \;\frac{\sqrt{L\,\ell_{z}}}{\xi}
  \\
  &\approx& \eta
  \;\sqrt{\frac{\ell_{xy}^2\,\ell_z}{\xi^3}}
  \;\sqrt{\frac{L}{\xi}}
  \;.
\end{eqnarray*}
where $\ell_{xy}\sim a$ is the correlation length of the disorder in
the plane,  $\ell_{z}$ the correlation length in the paraxial
direction, and $\xi$ is the horizontal extent of the zero mode.

\subsection{Adiabatic condition}

Braiding vortices in the vertical direction $z$ must be done slowly enough to satisfy the adiabatic condition. We introduce factors of $\hbar$ and define $t \equiv z/c$ to work in units more familiar from quantum mechanics. The amplitude to be in the $n^{\textrm{th}}$ mode, given that the state starts in
the zero mode, is given by
\begin{eqnarray*}
  c_n(t) &=&
  -\int_0^t dt'\; \frac{\bra{n}\dot{H}\ket{0}}{E_0-E_n} 
  \;
  e^{i(E_0-E_n)t'/\hbar}
  \;
  c_0(t')
  \\
  &=&
  -\hbar
  \frac{\bra{n}\dot{H}\ket{0}}{(E_0-E_n)^2}
  \; e^{i\frac{1}{2}(E_0-E_n)t}
  \;  2\,\sin\left(\frac{1}{2}(E_0-E_n)t\right)
  \;.
\end{eqnarray*}
One can estimate the probability $p_{\rm out}(t)=1-|c_0(t)|^2$ to
escape the zero mode. Using $1-|c_0(t)|^2 = \sum_{n \ne 0} |c_n(t)|^2$, we have
\begin{eqnarray*}
  p_{\rm out} &\approx&
    (2\hbar)^2
    \sum_{n\ne 0}
    \;\frac{\bra{0}\dot{H}\ket{n}\,\bra{n}\dot{H}\ket{0}}{(E_n-E_0)^4}
    \\
    &\le&
    \frac{(2\hbar)^2}{\Delta^4}
    \sum_{n \ne 0}
    \;\bra{0}\dot{H}\ket{n}\,\bra{n}\dot{H}\ket{0}
    \\
    &=&
    \frac{(2\hbar)^2}{\Delta^4}
    \;\left(\bra{0}\dot{H}^2\ket{0}-\bra{0}\dot{H}\ket{0}^2\right)
    \\
    &=&
    \frac{(2\hbar)^2}{\Delta^4}
    \;\bra{0}\dot{H}^2\ket{0}_c
    \;.
\end{eqnarray*}
Thus we arrive at
\begin{eqnarray*}
  p_{\rm out} &\lesssim&
    \frac{(2\hbar)^2}{\Delta^4}
    \;\bra{0}\dot{H}^2\ket{0}_c
    \;.
\end{eqnarray*}

In our case, say we move the vortex in space with velocity $\dot R$,
then
$\bra{0}\dot{H}^2\ket{0}_c=\dot{R}^2\,\bra{0}{H'}^2\ket{0}_c\approx
\dot{R}^2\,(\Delta/\xi)^2$, where $\xi$ is again the horizontal extent of the zero mode. Consider the time $\tau_\xi$ it takes to
move the vortex a distance $\xi$. In the same time, light propagates a
distance $L_\xi$ (in the $z$-direction) along the guided mode. Notice
that in the units that we use above, $\Delta$ is energy, not
wavenumber. From our tight-binding Hamiltonian we can estimate $\Delta \sim \hbar c / \xi$. We can then write
\begin{eqnarray*}
  p_{\rm out} &\lesssim&
    \frac{(2\hbar)^2}{\Delta^4}
    \;\left(\frac{\xi}{\tau_\xi}\right)^2
    \,\left(\frac{\Delta}{\xi}\right)^2
    \\
    &\approx&
    \frac{(2\hbar c)^2}{\Delta^4}
    \;\left(\frac{\xi}{L_\xi}\right)^2
    \,\left(\frac{\Delta}{\xi}\right)^2
    \;,
\end{eqnarray*}
or, finally,
\begin{eqnarray*}
  p_{\rm out} &\lesssim&
    \;\left(\frac{\xi}{  L_\xi}\right)^2
    \;.
\end{eqnarray*}
Therefore, the condition for adiabaticity is that the distance $L_\xi$
along the paraxial direction needed to move the vortex a distance
$\xi$ in the $xy$-plane must be large compared to $\xi $, {\it i.e.},
$L_\xi\gg \xi $.

\subsection{Number of braids}

The total length $L$ needed to do $N$ braids is $L\approx \pi
N\;L_\xi\,d/\xi$, where $d$ is the distance between the
vortices. (Basically, it takes $\pi d/\xi$ movements of the vortex by
a distance $\xi$, and adiabaticity requires that the distance in the
paraxial direction be $L_\xi$ for each move.)

The upper limit on the number of braids is reached when
$\delta\phi\sim \pi$, hence
\begin{eqnarray*}
  \pi\;\approx\;\delta\phi 
  &\approx& \eta
  \;\sqrt{\frac{\ell_{xy}^2\,\ell_z}{\xi^3}}
  \;\sqrt{\frac{L}{\xi}}
  \\
  &\approx& \eta
  \;\sqrt{\frac{\ell_{xy}^2\,\ell_z}{\xi^3}}
  \;\sqrt{\frac{L_\xi}{\xi}}
  \;\sqrt{\frac{d}{\xi}}
  \;\sqrt{\pi\,N}
    \;,
\end{eqnarray*}
leading to
\begin{eqnarray*}
  N &\approx& \pi\;{\eta}^{-2}
  \;\frac{\xi}{d}
  \;\frac{\xi}{L_\xi}
  \;\frac{\xi^3}{\ell_{xy}^2\,\ell_z}
  \;.
\end{eqnarray*}
Now, the ratio $n_d=d/\xi\sim 20$ guarantees that the zero modes in
different vortices do not mix. The ratio $ \xi /L_\xi=\sqrt{p_0}$, and
we want to keep $p_0\sim 0.01$, say. So we can write
\begin{eqnarray*}
  N &\approx& \pi\;{\eta}^{-2}
  \;\frac{\xi}{d}
  \;\frac{\xi}{L_\xi}
  \;\frac{\xi^3}{\ell_{xy}^2\,\ell_z}
  \\
  &\approx& \pi
  \frac{\sqrt{p_o}}{\eta^2 \,n_d}
  \;\frac{\xi^3}{\ell_{xy}^2\,\ell_z}
  \;.
\end{eqnarray*}
A few comments are in order. First, notice that, as expected, if there
is no disorder ($\eta\to 0$), the number of braids is
unbounded. Second, even if $\eta$ is non-zero, there is a controlled
limit where the number of braids can be made as large as desired: one
can take the limit of large vortex sizes $\xi$. 

Say one takes a pessimistic estimation using $\eta\sim 5\%$, and
$\xi\sim 5a$. Then, $N\approx \pi/.5\times\;5^3\approx 800$. For
$\eta\sim 1\%$ and $\xi\sim 10a$, $N\approx \pi/.02\times\;10^3\approx
160,000$. And for $\eta\sim 1\%$ and $\xi\sim 50a$, $N\approx
\pi/.02\times\;50^3\approx 20,000,000$.

One can also estimate the vertical length needed per braid, $L/N = \pi n_d \xi / \sqrt{p_0}$. Taking $a \sim 10 \, \mu$m and all other estimates as above, we have $L/N \approx 3$ cm if $\xi \sim 5a$ and $L/N \approx 6$ cm if $\xi \sim 10a$.

Finally, it is noteworthy that the correlation length in the paraxial
direction is not necessarily of the order of $a$, it can be {\it much
  smaller}. This is because it comes from fluctuations in writing the
waveguides during their growth. So $\ell_z$ can be possibly hundredths
of times smaller than $a$, in which case the number of braids
estimated above goes up by a factor of 100.

\subsection{Finite size effects}

In a lattice with edges, each vortex will create both a mode localized at the vortex and a mode localized at the edge of the system. If the lattice is small enough, there will be an ``energy" splitting between symmetric and antisymmetric linear combinations of our zero mode and edge mode. Let us assume the splitting to be proportional to $t\, e^{- \frac{\Delta d_E}{t a}}$. To prevent the zero mode from tunneling to the edge (a distance $d_E$ away) during the course of braiding, we require

\begin{equation}
L \ll \frac{\pi}{t} e^{ \frac{\Delta d_E}{t a}}.
\end{equation}

\section{Beam splitters and interference of coherent states}

In this Appendix, we elaborate on the use of beam splitters to interfere the coherent states of light used in our experimental proposal.  A beam splitter can be thought of as an optical circuit element that interferes photons entering two input channels and outputs the result to two output channels.  At the level of single photons, a 50/50 dielectric beam splitter acts as follows:
\begin{equation}
\left(\begin{matrix}
\hat b^\dagger_1\\
\hat b^\dagger_2
\end{matrix}\right)\longrightarrow
\frac{1}{\sqrt 2}\left(\begin{matrix}
1& -i\\
-i & 1
\end{matrix}\right)
\left(\begin{matrix}
\hat b^\dagger_1\\
\hat b^\dagger_2
\end{matrix}\right)=
\frac{1}{\sqrt 2}\left(\begin{matrix}
\hat b^\dagger_1-i\, \hat b^\dagger_2\\
-i\, \hat b^\dagger_1+\hat b^\dagger_2
\end{matrix}\right).
\end{equation}
The factors of $i$ above arise from $\pi/2$ phase shifts of the photons in each channel upon reflection within the beam splitter.  This action on single photons in turn induces an action on coherent states of photons:
\begin{equation}
\hspace{-3in}|\lambda_1,\lambda_2\rangle=e^{-(|\lambda_1|^2+|\lambda_2|^2)/2}\, e^{\lambda_1\, \hat{b}^\dagger_1}\, e^{\lambda_2\, \hat{b}^\dagger_2}\, |0,0\rangle\nonumber
\end{equation}
\begin{equation}
\qquad\qquad\longrightarrow e^{-(|\lambda_1|^2+|\lambda_2|^2)/2}\, e^{\lambda_1(\hat{b}^\dagger_1-i\, \hat{b}^\dagger_2)/\sqrt{2}}\, e^{\lambda_2(-i\, \hat{b}^\dagger_1+\hat{b}^\dagger_2)/\sqrt{2}}\, |0,0\rangle=\left |\frac{\lambda_1-i\, \lambda_2}{\sqrt{2}},\frac{-i\, \lambda_1+\lambda_2}{\sqrt{2}}\right\rangle.
\end{equation}
Note that the output coherent state is not a superposition of coherent states, but rather a new coherent state with a 50/50 admixture of photons from the two channels.

The interferometer proposed in the main text uses two sets of three beam splitters, one for each vortex core: the first splits each incoming laser beam into two equal parts, and the second interferes light from the two branches of the interferometer.  Let us now follow one of the laser beams through each optical circuit element.  We denote the initial state of light from one of the three laser beams by $|\lambda^\prime_j,0\rangle$.  The effect of the first beam splitter is to split this beam in two:
\begin{equation}
|\lambda^\prime_j,0\rangle\longrightarrow\left |\frac{\lambda^\prime_j}{\sqrt 2},\frac{-i\, \lambda^\prime_j}{\sqrt 2}\right\rangle\equiv |\lambda_j,\tilde\lambda_j \rangle.
\end{equation}
From here, the light in the coherent state $|\lambda_j\rangle$ enters the waveguide lattice executing the braid $\hat B_1\hat B_2$, while the light in the coherent state $|\tilde\lambda_j\rangle$ enters the waveguide lattice executing the braid $\hat B_2\hat B_1$.  The state of light entering the second beamsplitter is then $|\pm \lambda_j,\pm\tilde\lambda_j\rangle$, where the choice of sign on each coherent state depends on the braid that was performed on that state.  Passing this state through the beamsplitter gives
\begin{align}
|\pm \lambda_j,\pm\tilde\lambda_j\rangle\longrightarrow\left|\frac{\pm \lambda_j\mp i\, \tilde \lambda_j}{\sqrt 2},\frac{\mp i\, \lambda_j\pm \tilde\lambda_j}{\sqrt 2}\right\rangle=\left|\frac{\pm \lambda_j\mp \lambda_j}{\sqrt 2},\frac{-i(\pm \lambda_j\pm\lambda_j)}{\sqrt 2}\right\rangle.
\end{align}
Therefore, light exiting one output channel of the beam splitter experiences destructive interference, while light exiting the other channel experiences constructive interference.  Whether this occurs in the first or second output channel depends on the relative sign between $\lambda_j$ and $\tilde\lambda_j$ after each beam exits its respective waveguide lattice.

\hfill

\end{widetext}

\end{document}